\documentclass{article}

\usepackage{PRIMEarxiv}

\usepackage[utf8]{inputenc} 
\usepackage[T1]{fontenc}    
\usepackage{hyperref}       
\usepackage{url}            
\usepackage{booktabs}       
\usepackage{amsfonts}       
\usepackage{nicefrac}       
\usepackage{microtype}      
\usepackage{lipsum}
\usepackage[numbers]{natbib}
\usepackage{fancyhdr}       
\usepackage{graphicx}       
\usepackage[export]{adjustbox}
\usepackage{siunitx}
\usepackage{todonotes}
\usepackage{caption}
\usepackage{subcaption}
\graphicspath{{media/}}     
\usepackage{tikz,xcolor,hyperref}
\definecolor{lime}{HTML}{A6CE39}
\DeclareRobustCommand{\orcidicon}{%
    \begin{tikzpicture}
    \draw[lime, fill=lime] (0,0) 
    circle [radius=0.16] 
    node[white] {{\fontfamily{qag}\selectfont \tiny ID}};    \draw[white, fill=white] (-0.0625,0.095) 
    circle [radius=0.007];    \end{tikzpicture}
    \hspace{-2mm}}
\foreach \x in {A, ..., Z}{%
    \expandafter\xdef\csname orcid\x\endcsname{\noexpand\href{https://orcid.org/\csname orcidauthor\x\endcsname}{\noexpand\orcidicon}}
    }
\title{Collective bistability of pyridine-furan nanosprings coupled by a graphene plate
}

\author{
  Maria A. Frolkina$^{1,2}$\orcidD, Anastasia A. Markina$^{1,2,c}$\orcidA, Vladislav S. Petrovskii$^{1,2}$\orcidF, Alexey M. Astakhov$^{2}${},\\ 
  \textbf{Alexander D. Muratov$^{1,2}$\orcidE, Alexander F. Valov$^{2}$ and Vladik A. Avetisov$^{1,2}$\orcidB } \\
  $^{1}$ Semenov Federal Research Center for Chemical Physics, Russian Academy of Sciences, Moscow, Russia \\
  $^{2}$ Design Center for Molecular Machines, Moscow, Russia \\
  \texttt{$^{c}$ dr.anastasia.markina@gmail.com} \\
}

\begin{document}
\maketitle

\begin{abstract}
\end{abstract}
Nanometer-sized molecular structures exhibiting mechanical-like switching between discrete states are of great interest for their potential uses in nanotechnology and materials science. Designing such structures and understanding how they can be combined to operate synchronously is a key to creating nanoscale functional units. Notable examples recently discovered using atomistic simulations are pyridine-furan and pyridine-pyrrole nanosprings. When slightly stretched in aqueous or organic solutions, these nanosprings exhibit bistable dynamics akin to Duffing nonlinear oscillators. Based on these findings, we designed a hybrid system consisting of several pyridine-furan nanosprings attached to a graphene plate in organic solvent and simulated the molecular dynamics of the construct. Our focus is on how the nanosprings coupled by a graphene plate work together, and whether such a design enables the nanosprings to respond synchronously to random perturbations and weak external stimuli. Molecular dynamics simulations of this specific construct are complemented by a theoretical model of coupled bistable systems to understand how the synchronization depends on coupling of bistable units.    

\keywords{Nanosprings \and nonlinear dynamics \and bistability \and spontaneous vibrations \and stochastic resonance \and Duffing oscillators}

\section{Introduction}
The use of nanometer-sized molecules as nonlinear dynamic systems capable of abrupt transitions between discrete states is of significant interest in the development of operational units with the characteristics of nanomachines. These systems can serve as switches and logic gates \cite{mi6081046,C5SC02317C,C7CS00491E,benda_substrate-dependent_2019,berselli_robust_2021,NICOLI2021213589}, sensors and actuators \cite{zhang_molecular_2018,shu_stimuli-responsive_2020,lemme_nanoelectromechanical_2020,shi_driving_2020,aprahamian_future_2020}, and mechanoelectrical transducers and energy harvesters \cite{li_energy_2014,kim_harvesting_2015,ackerman_anomalous_2016,dutreix_two-level_2020,thibado_fluctuation-induced_2020}. In addition to practical applications, the dynamics of nanometer-sized systems are currently attractive for substantiating basic concepts of stochastic thermodynamics underlying the operation of molecular machines \cite{evans_fluctuation_2002,seifert_stochastic_2012,horowitz_thermodynamic_2020,ciliberto_experiments_2017}.

Despite the clear potential of nonlinear dynamical systems operating as nanometer-scale machines, their practical implementation is hampered by the lack of guidance regarding target molecular structures. Atomistic modeling appears to be a very effective technique for providing these insights. In particular, our recent studies \cite{avetisov2019oligomeric,avetisov2021short,astakhov_spontaneous_2024,markina_spontaneous_2023} have highlighted the potential of using molecular dynamics simulations to design specific examples of nanometer-sized oligomers that behave as bistable dynamical systems, drawing comparisons with classical mechanical constructs such as Euler arches and Duffing springs. These oligomers provide a rich basis for understanding the distinctions in the bistable dynamics of nanometer-scale systems compared to larger systems. For example, a deterministic force applied to an oligomeric Euler arch or an oligomeric Duffing oscillator can dictate switching between two states, allowing the oligomeric system's response to be controlled by passing through a critical value of the force. In this respect, bistable oligomers behave similarly to classical bistable constructs. However, near the critical force, bistable oligomers exhibit spontaneous vibrations between the two states due to random impacts from solvent molecules. According to theory \cite{Gammaitoni1998}, spontaneous vibrations are observed when the intensity of random perturbations is sufficient to induce transitions over the bistability barrier separating the two energy minima. Comparability of the bistability barrier with the mean energy of random perturbations is essential for the occurrence of spontaneous vibrations. The bistability barriers of macroscopic-sized Euler arches and Duffing springs, even as small as a micron, are much higher than the mean energy of thermal fluctuations, necessitating much stronger perturbations to activate their spontaneous vibrations. In contrast, the bistability barriers of nanoscale systems can be reasonably low, allowing spontaneous vibrations to be activated by ordinary thermal fluctuations while still being high enough to separate the two states against thermal fluctuations. The ratio found for oligomeric Euler arches and Duffing oscillators \cite{avetisov2019oligomeric,avetisov2021short} can serve as a benchmark.

Notable examples of such oligomers are pyridine-furan and pyridine-pyrrole springs a few nanometers in size \cite{avetisov2019oligomeric}, which act as bistable Duffing oscillators. Up to a certain critical stretch, these nanosprings behave like ordinary springs with linear elasticity. However, beyond the critical stretch, the nanosprings become bistable and switch between two states when the stretching force exceeds a critical value. Near the critical stretch, they exhibit spontaneous vibrations at MHz frequencies activated by room-temperature thermal fluctuations of the surrounding solvent. Remarkably, the spontaneous vibrations of the nanosprings are highly sensitive to weak oscillatory stimuli, demonstrating the stochastic resonance effect.

Furthermore, coupling two pyridine-furan nanosprings with a rigid oligomeric bridge can lead to spontaneous synchronization of their vibrations \cite{avetisov2021short}, turning the entire construction into a larger bistable system. Although coupling two nanosprings with an oligomeric bridge did not achieve full synchronization, it showed that collective bistability can be achieved constructively, supporting the idea of using spontaneous synchronization to scale the bistable dynamics of individual nanosprings to larger systems. The question remains: what design can ensure the synchronization of spontaneous vibrations among a reasonably large number of nanosprings?

In this article, we present an atomistic model of such a design, reflecting the idea of hybrid systems with nanosprings placed between a fixed substrate and a moving plate. The substrate positions the nanosprings regularly, while the moving plate covers the nanosprings and can fluctuate “up and down” due to their spontaneous vibrations. The hybrid system resembles a 2-D metamaterial with thickness determined by the size of incorporated nanosprings, while the material area can be scaled by the number of nanosprings. By changing this number, it seems possible to scale the hybrid system to submicron and micron sizes, maintaining the unique bistable dynamics of nanoscale oligomeric springs, such as thermally activated bistability and sensitivity to weak external stimuli.

Intuitively, the moving plate in this design must be rigid to couple the nanosprings and force them to vibrate synchronously. However, even if the moving plate is rigid enough to ensure synchronous vibrations, it is unclear whether a large number of nanosprings can vibrate under such constraints. It is also unclear what characteristics of collective bistability should be expected in such a design. We explore this question from two perspectives. 

First, we analyze a theoretical model of coupled bistable systems to provide qualitative guidelines on how synchronization depends on the number of mutually coupled bistable systems and their coupling strength. Interestingly, contrary to the intuitive desire to make the moving plate as rigid as possible, the theoretical analysis suggests a contrary approach to varying the coupling of bistable systems to achieve high synchronization. Second, we design an atomistic model of a hybrid system with seven pyridine-furan nanosprings attached to a movable graphene plate in tetrahydrofuran and study its molecular dynamics. We explore the synchronization of spontaneous vibrations of nanosprings in this hybrid system and show that the spontaneous vibrations of nanosprings in this hybrid system are quite synchronized. The vibrations of the graphene plate caused by the spontaneous vibrations of nanosprings are similar to the springs' behaviour, except for a shift to lower frequencies.

The following sections describe our methodology, present simulation results, and discuss their implications. We conclude with insights into the potential of these systems for future technological advancements and theoretical exploration.

\section{Matherials and Methods}
\label{sec:MatMet}

\subsection{Pyridine-Furan Springs affixed to a hexagonal graphene plate }\label{Sec:MatMet1}

\begin{figure}
    \centering
    \includegraphics[width=.95\textwidth]{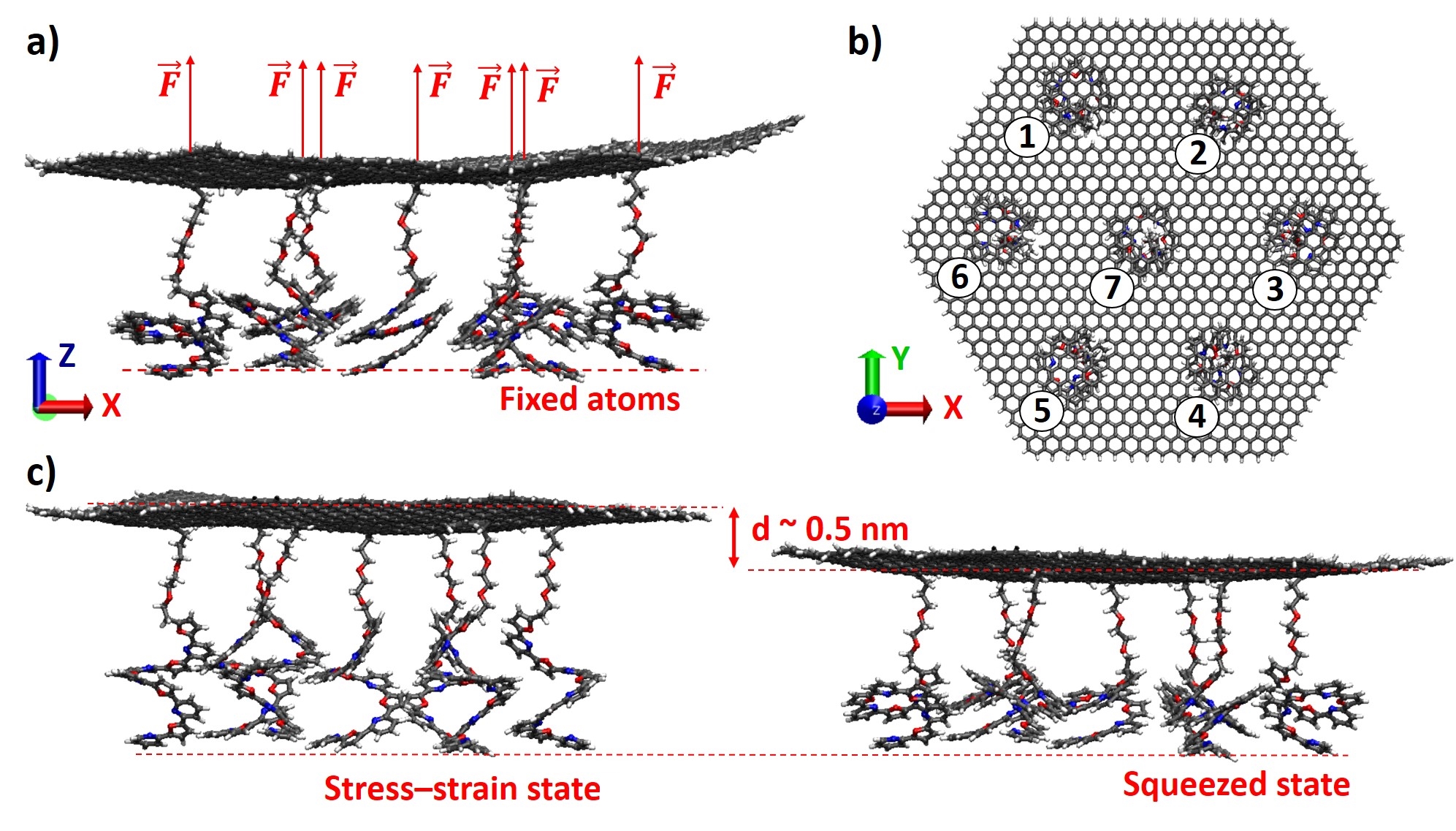}
    \caption{Computational model and deformation states of oligo-PF-5 nanosprings attached to a graphene plate. a) Side view illustrating seven oligo-PF-5 nanosprings attached to a graphene plate with applied longitudinal forces (F); b) Top view of the graphene plate showing the arrangement of nanosprings; c) Side view of the stress-strain and squeezed states of a nanosprings.}
    \label{fig:fig1}
\end{figure}

The pyridine furan (PF) oligomer is a conductive conjugated molecule composed of 5- and 6 member hetherocycles, incuding oxygen and nytrogen, respectively, as synthesized and characterized by \citet{jones1997extended}. These copolymers tend to adopt helix-like structures, stabilized by the $\pi-\pi$ interactions between aromatic groups located at adjacent turns \cite{sahu2015}. Such interactions are likely to contribute to the nonlinear elasticity observed in the oligo-PF-$5$ springs. To demonstrate the effect of synchronization, seven oligo-PF-5 nanosprings were affixed to a hexagonal graphene plate. The attachment was facilitated through a spacer composed of two ethylene glycol monomer units $-(C_2H_4O)_2-$, ensuring robust chemical bonding to the graphene substrate. The choice of graphene was strategic, providing a rigid backbone to the system, allowing it to operate cohesively during the synchronization of the springs. For detailed illustrations of the setup, refer to Figure \ref{fig:fig1}.  

This Figure \ref{fig:fig1} illustrates the computational models used in the study of oligo-PF-5 nanosprings attached to a hexagonal graphene plate under varying mechanical loads. Panel (a) of Figure \ref{fig:fig1} shows a side view of the system where longitudinal forces, denoted by $\vec{F}$, are applied to the upper ends of the nanosprings, simulating tensile stress conditions. To stabilize the configuration in the simulation box, the free-from-spacer ends of the springs were anchored.These fixed atoms at the base of the springs ensure stability during simulation (shown as red dashed lines). Panel (b) of Figure \ref{fig:fig1} provides a top view of the graphene plate, highlighting the symmetrical placement of seven nanosprings, enhancing the potential for synchronized vibrational behavior under stress.

Panel (c) of Figure \ref{fig:fig1} depicts the two primary deformation states of nanosprings under stress: the stress-strain state and the squeezed state, respectively. In the stress-strain state shown in panel (c) on the left, the nanospring is elongated, demonstrating the extension due to the applied force. Conversely, the right part of the panel (c) shows the squeezed state where the nanospring compresses, illustrating the potential energy storage within the system. 

Previous studies \cite{avetisov2019oligomeric,avetisov2021short} have shown that the addition of a charge and a fixed counterion can significantly alter the symmetric force parameters for bistability. Consequently, we incorporated unit charges into the graphene atoms at the points of attachment of each nanospring with a spacer. A tensile force was then applied to examine the system's behavior under charged conditions. To balance these charges, compensating counterions were positioned at a distance of 4 nm above the graphene plate, providing a stable electrostatic environment for studying spontaneous vibrations (SV) and stochastic resonance (SR) in this charged setup. Graphene and springs were solvated in a hydrophobic solvent, tetrahydrofuran (THF). The inter-turn distance was found to be closed to $\SI{0.35}{\nano\metre}$ in all non-stretched samples, consistent with the findings reported by \citet{sahu2015}.

\subsection{Simulation Details}\label{Sec:MatMet2}

The graphene and oligo-PF-springs, along with the environmental solvent, were modeled using a fully atomistic approach within a canonical (NVT) ensemble. The simulations were conducted in a box measuring $\SI{10.5}{} \times \SI{9.6}{} \times \SI{8.8}{\nano\metre\cubed}$, utilizing a timestep of $\SI{2}{\femto\second}$. We employed Gromacs 2023 \cite{abr2015} coupled with the OPLS-AA force field \cite{kam2001} (for additional simulation details, refer to our previous publications \cite{avetisov2019oligomeric,avetisov2021short,astakhov_spontaneous_2024,markina_spontaneous_2023}). The temperature was maintained at $\SI{320}{\kelvin}$, corresponding to the equilibrium state of the springs \cite{sahu2015}, controlled by the Nosé–Hoover thermostat. Each dynamic trajectory lasted $\SI{1}{\micro\second}$ and was repeated multiple times to enhance statistical reliability. The system was equilibrated in the NPT ensemble with a timestep of 0.01 fs at a temperature of 320 K until the system volume stabilized, which typically required 150-200 ps. During equilibration, the velocity-rescaling thermostat and the Berendsen barostat were employed.

To study system dynamics, a tensile force ($\vec{F}$) was uniformly applied along the spring axis to each carbon atom on the graphene plate where a spring with a spacer is attached (see Figure 1). Under conditions of mild tensile stress, the springs maintained their initial compressed state due to stacking interactions, exhibiting only slight extensions in accordance with linear elasticity principles. Upon reaching a ${F_s}$ of approximately $\SI{103}{\pico\newton}$, each nanospring entered a mode of spontaneous vibration, a force value closely aligning with the critical threshold for a single oligo-PF-5 spring (see Supporting Information). Visualizations of complete and partial synchronization among the springs are available in the Supporting Information.

The variable $R_{e}$, representing the distance between the ends of a spring (excluding the spacer molecule), was used as a collective descriptor of the spring's long-term dynamics. Bistability was clearly defined with two reproducibly distinct states characterized by end-to-end distances of $R_{e} \sim \SI{1.10}{\nano\metre}$ and $R_{e} \sim \SI{1.45}{\nano\metre}$, referred to as the squeezed and stress-strain states, respectively. Synchronization among the springs was quantitatively evaluated by constructing a correlation matrix, where each element denotes the Pearson correlation coefficient between the spontaneous vibration trajectories of each spring pair.

\section{Results}

\subsection{Theoretical modelling: Langevin approximation}\label{Sec:MatMet0}
Here, we introduce the theoretical underpinning of the bistable system's synchronization to bridge the molecular dynamics simulations presented below with the concept of synchronization. Let us consider the overdamped regime of $N$ coupled identical Duffing oscillators first discussed by Neiman for $N=2$ in \cite{neiman_synchronizationlike_1994}. The $N$-dimensional dynamics of the system in dimensionless units is described by Langevin equations:
\begin{equation}
\frac{dx_i}{dt}=-\frac{dU(x,t)}{dx}-\sum\limits_{i\neq j}^{N}A_{ij}\frac{d \Phi(x_i,x_j)}{dx_i}+\sigma\xi_i(t),
\label{sde}
\end{equation}
where $x_i$ ($i=\overline{1,N}$) is the coordinate of the $i$-th Duffing oscillator, $\xi_i(t)$ is a delta-correlated white noise, $\sigma$ is a noise amplitude, and \(U(x)=-\frac{\alpha}{2}x^2+\frac{1}{4}x^4 \) is a double-well potential. The parameter $\alpha$  determines the positions of the potential minima, $x_{min}=\pm\sqrt{\alpha}$, and the bistability barrier, $\Delta U=\alpha^2/4$, separating the minima. The symmetric matrix $A$, also referred to as an adjacency matrix, represents the topology of particle-particle interactions, i.e., $A_{ij}=1$ if particles interact and $A_{ij}=0$ otherwise. Particles interact via harmonic potential $\Phi(x,y)$:
\begin{equation}
\Phi(x,y)=\frac{b}{2}(x-y)^2,
\label{CF}
\end{equation}
where $b>0$ is the coupling constant.

We will solve the equations (\ref{sde}) for $N=7$ particles using Heun's method to represent the synchronization effect. Specifically, we choose dimensionless parameters $\alpha=10$, $\sigma^2=18$ for which the Kramers rate $r_K\propto\exp\left(-\frac{\alpha^2}{2\sigma^2}\right)$ is close to that obtained in molecular simulations. Particles interact via $7\times 7$ adjacency matrix $A$
\begin{equation}
    A=\left(
\begin{array}{ccccccc}
 0 & 1 & 0 & 0 & 0 & 1 & 1 \\
 1 & 0 & 1 & 0 & 0 & 0 & 1 \\
 0 & 1 & 0 & 1 & 0 & 0 & 1 \\
 0 & 0 & 1 & 0 & 1 & 0 & 1 \\
 0 & 0 & 0 & 1 & 0 & 1 & 1 \\
 1 & 0 & 0 & 0 & 1 & 0 & 1 \\
 1 & 1 & 1 & 1 & 1 & 1 & 0 \\
\end{array}
\right),
\label{adj}
\end{equation}
whose graph representation is a hexagone with a node in the center, it is shown in Figure S2.

Before proceeding to the results of numerical simulations, let us provide some qualitative arguments on the collective behavior of the system. In general, it depends on the interplay between interparticle interactions $\Phi(x,y)$, their topology $A$, confining potential $U(x,t)$, and noise amplitude $\sigma$. Let us define the critical value of the coupling constant, $b_{cr}$, at which interparticle interactions acting on a particle balance the bistability barrier separating potential wells:
\begin{equation}
\frac{b_{cr}}{2}\sum\limits_{i=1}^{N_{\text{nn}}}(2 x_{\text{min}})^2 =\Delta U,
\end{equation} 
where $N_{\text{nn}}$ is the average number of nearest neighbors of the graph defined by the adjacency matrix $A$. After some algebra, we obtain the critical value of the coupling constant
\begin{equation}
b_{cr}=\frac{\alpha}{8N_{nn}},
\label{est}
\end{equation}
that separates two different regimes: (i) Stochastic regime with $b\ll b_{cr}$ -  where interparticle interactions effectively lowers the bistability barrier, which increases Kramers rate and hence, reduces the average time of the interwell transitions; and (ii) Frozen regime with $b\gg b_{cr}$ -  where interparticle forces drag  an escaping particle back, not allowing for interwell transitions.

\begin{figure}[htbp]
    \centering
    \begin{subfigure}{0.48\textwidth}
        \centering
        \includegraphics[width=\textwidth,valign=c]{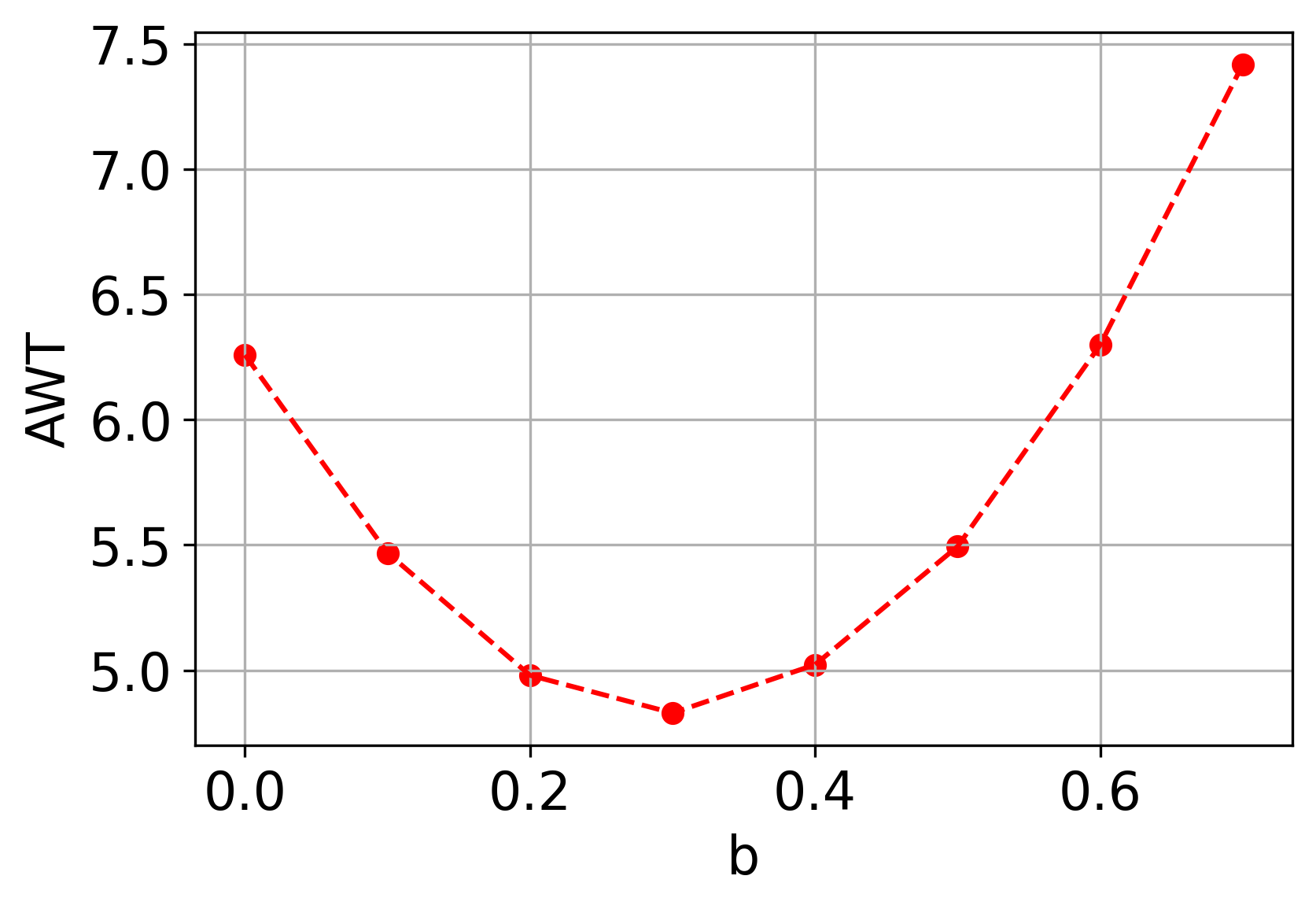}
        \label{fig:Fig_res_a}
        \caption{}
    \end{subfigure}
    \hfill 
    \begin{subfigure}{0.48\textwidth}
        \centering
        \includegraphics[width=\textwidth]{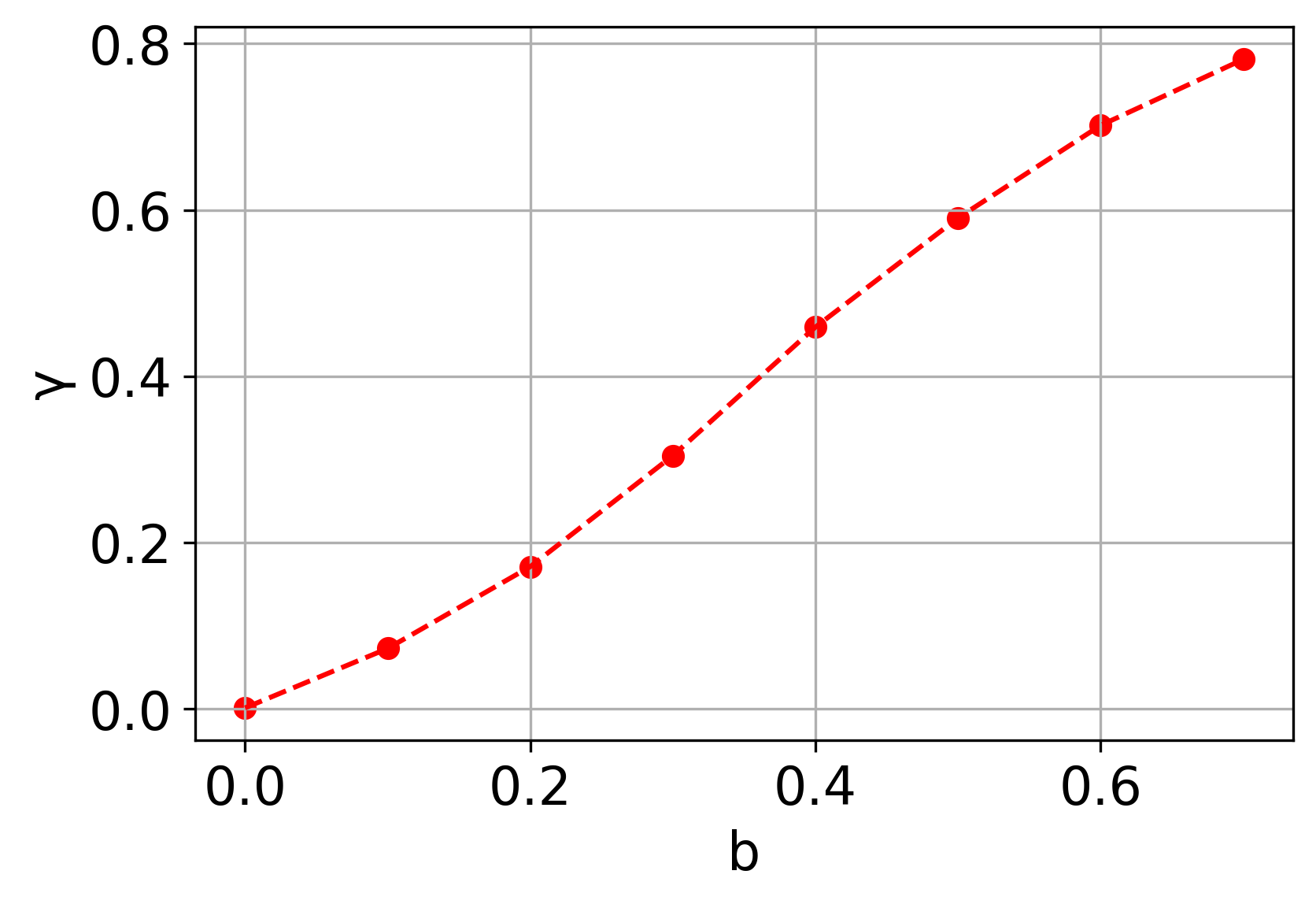}
        \caption{}
        \label{fig:Fig_res_b}
    \end{subfigure}
    \caption{(a) Average waiting time as a function of coupling constant $b$. (b) The system-averaged correlation coefficient $\gamma$ as a function of coupling constant $b$.}
    \label{fig:Fig_res}
\end{figure}

We characterize the dynamical behavior of the system using two system-averaged quantities: (i) average waiting time (AWT) - an arithmetic mean of the time that particles spend near potential minima $-x_{\text{min}}$ and $x_{\text{min}}$; and (ii) the averaged pairwise correlation coefficient 
\begin{equation}
\langle \gamma \rangle =\frac{1}{N(N-1)}\sum\limits_{i<j}\frac{\text{cov}(x_i(t),x_j(t))}{\sqrt{\text{cov}(x_i(t),x_i(t))}\sqrt{\text{cov}(x_j(t),x_j(t))}}
\end{equation}
between numerically generated tracjectories $x_i(t)$.  Figure (\ref{fig:Fig_res}) presents the dependence of the AWT and correlations on coupling constant $b$ for $N=7$ coupled via adjacency matrix $A$ (\ref{adj}) Duffing oscillators (\ref{sde}). One can see that the AWT  has a minimum at $b_s\approx 0.3$, separating two distinct collective behaviors described above: (i) stochastic  ($b<b_s$), where a significant reduction of the AWT occurs and (ii) frozen regime ($b>b_s$), where AWT rapidly grows. The qualitative estimate (\ref{est}) of the critical value of coupling constant  $b_{cr}\approx 0.36$ is in good agreement with the results of the numerical simulations. The averaged correlation coefficient, on the contrary, grows monotonically, with an inflection point at $b_s$ separating the system's behavior regimes with low and high pair correlation coefficients. 

\label{sec:Res}
\subsection{Spontaneous vibrations in the system of seven nanosprings
}\label{sec:Res1}

\begin{figure}[htbp]
    \centering
    \begin{subfigure}{0.48\textwidth}
        \centering
        \includegraphics[width=\textwidth,valign=c]{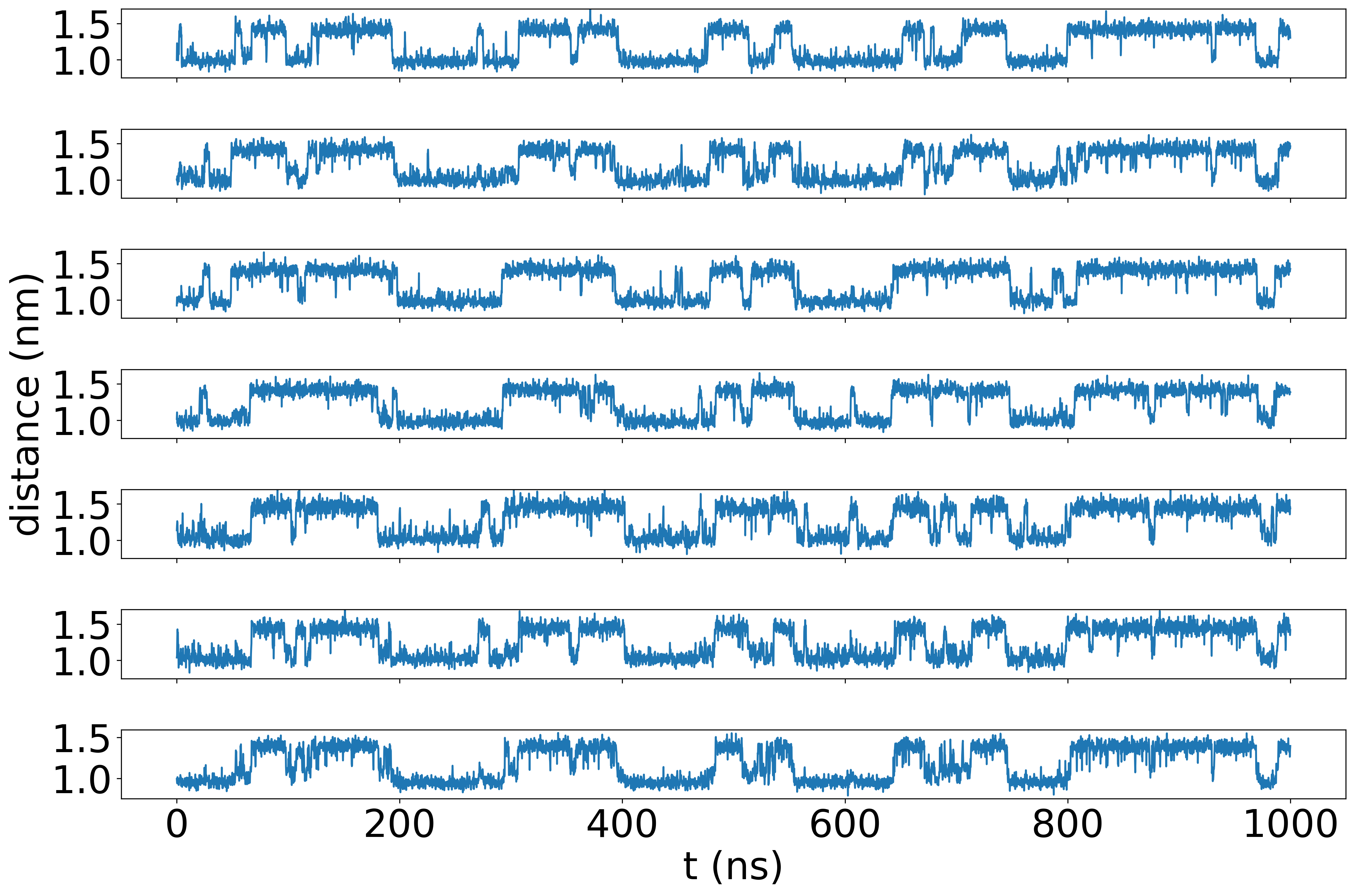}
        \label{fig:figSVa}
        \caption{}
    \end{subfigure}
    \hfill 
    \begin{subfigure}{0.48\textwidth}
        \centering
        \includegraphics[width=\textwidth]{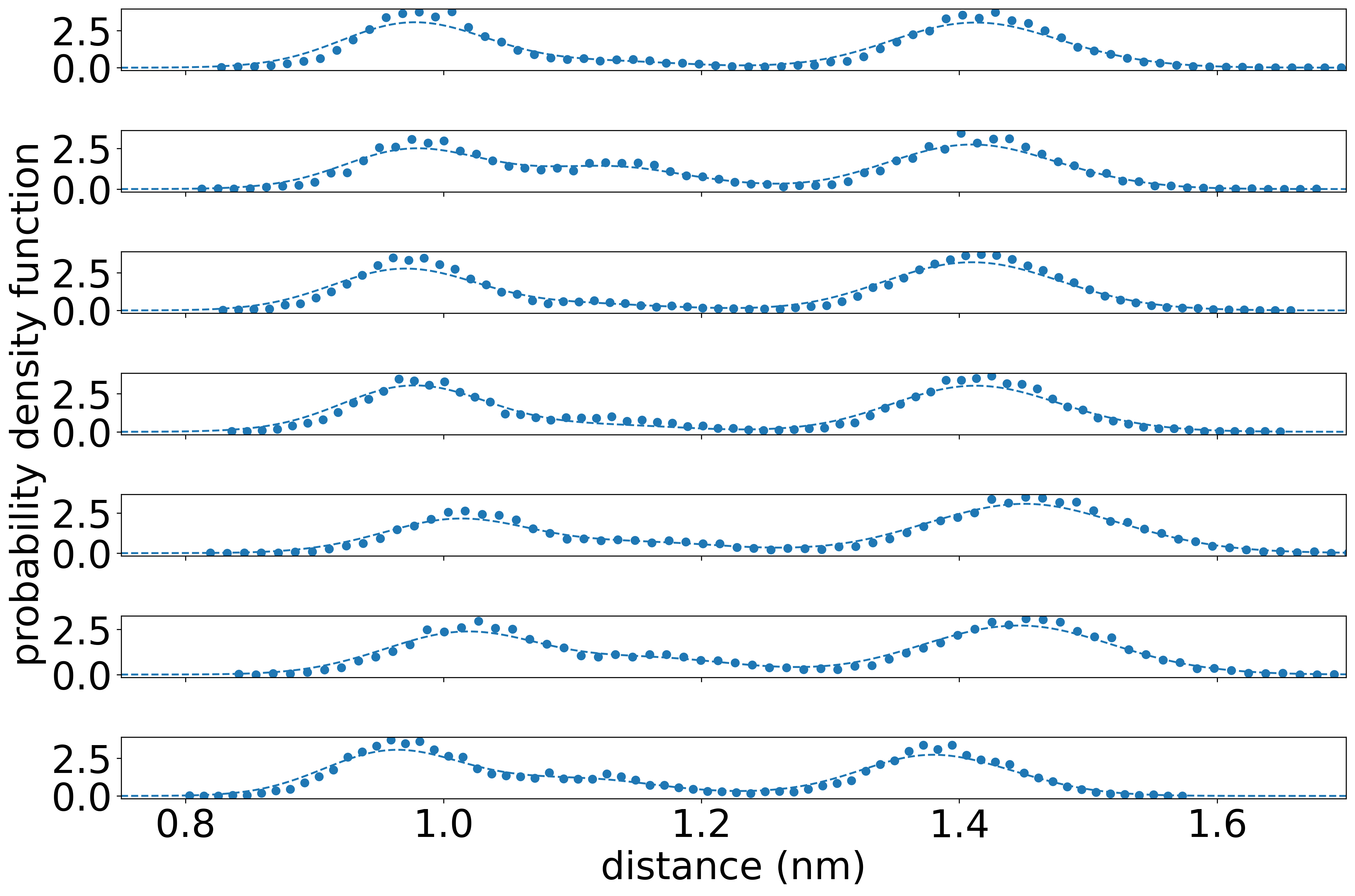}
        \caption{}
        \label{fig:figSVb}
    \end{subfigure}
    \caption{Spontaneous vibrations of 7 oligo-PF-5 nanosprings.(a) Temporal dynamics of end-to-end distances in seven oligo-PF-5 nanosprings under spontaneous vibrations. Each panel represents the trajectory of distance changes between the ends of one spring subjected to a tensile force \(F \approx \SI{103}{\pico\newton}\) over a 1000 ns simulation period. (b) Probability density functions for the end-to-end distance of seven oligo-PF-5 nanosprings under equilibrium conditions. Each panel represents the distribution of distances over a 1000 ns simulation period, illustrating the probability of each spring.}
    \label{fig:figSV}
\end{figure}

Figure \ref{fig:figSV}a presents the temporal dynamics of the end-to-end distances for seven oligo-PF-5 nanosprings experiencing spontaneous vibrations when subjected to a critical tensile force. Each graph displays the fluctuations in distance over a simulation time of $\SI{1}{\micro\second}$, illustrating how each spring responds uniquely to the imposed mechanical stress. The plots reveal a range of behaviors, with some springs exhibiting more pronounced vibrations, while others display relatively stable behavior. These differences are indicative of the complex interplay between mechanical forces and the molecular architecture of each nanospring. This variability in response under identical force conditions underscores the nuanced nature of nanoscale material behavior and highlights the importance of considering individual structural dynamics in the design and analysis of nanoscale mechanical systems.

Figure \ref{fig:figSV}b illustrates the probability density functions of the end-to-end distances for seven individual oligo-PF-5 nanosprings, derived from the state visits over the course of a $\SI{1}{\micro\second}$ simulation at the symetrical force $F_{s}$ condition. This symetrical force is a longitudinal load that result in equal visits of two states of vibrating springs. The panels depict how the end-to-end distances of each nanospring vary, with the x-axis representing the distance and the y-axis the probability density of these distances. Notably, each histogram shows a superposition of two well-defined bell-shaped curves. This indicates normal distributions around two mean values, which suggests that the springs spend most of their time in these highly probable states.

In the region of symetrical force $F_{s}$, the spontaneous vibrations of the oligo-PF-$5$ spring are most pronounced and symetric. Clear vibrations of the spring can be observed without any additional random perturbations applied to activate them. Instead, these vibrations are solely activated by thermal-bath fluctuations. In the symmetrical bistability region, neither the squeezed state nor the stress-strain state dominates, resulting in the mean lifetimes of the two states being approximately equal to $\tau=\SI{50}{\nano\second}$. In the case of a single spring, the mean lifetime is significantly shorter due to the lack of connectivity within the system, which results in less stable and shorter-lived bistable states compared to the interconnected nanospring array. The lack of interaction with other springs in the single spring setup results in quicker transitions between states, reducing the overall stability and duration of each state. 

\subsection{Stochastic resonance in the system of seven nanosprings
}\label{sec:Res2}

To investigate the stochastic resonance mode of the oligo-PF-$5$ spring, we introduced an additional weak oscillating force applied to the pulling end of the spring. This oscillating force was modeled by applying an oscillating electrical field, $E=E_{0} \cos (2\pi \nu t)$, to a unit charge located on the pulling end of the spring (for more details, refer to the "Parameters of periodic signal" section in the Supporting Information). 

\begin{figure}[htbp]
    \centering
    \begin{subfigure}{0.48\textwidth}
        \centering
        \includegraphics[width=\textwidth,valign=c]{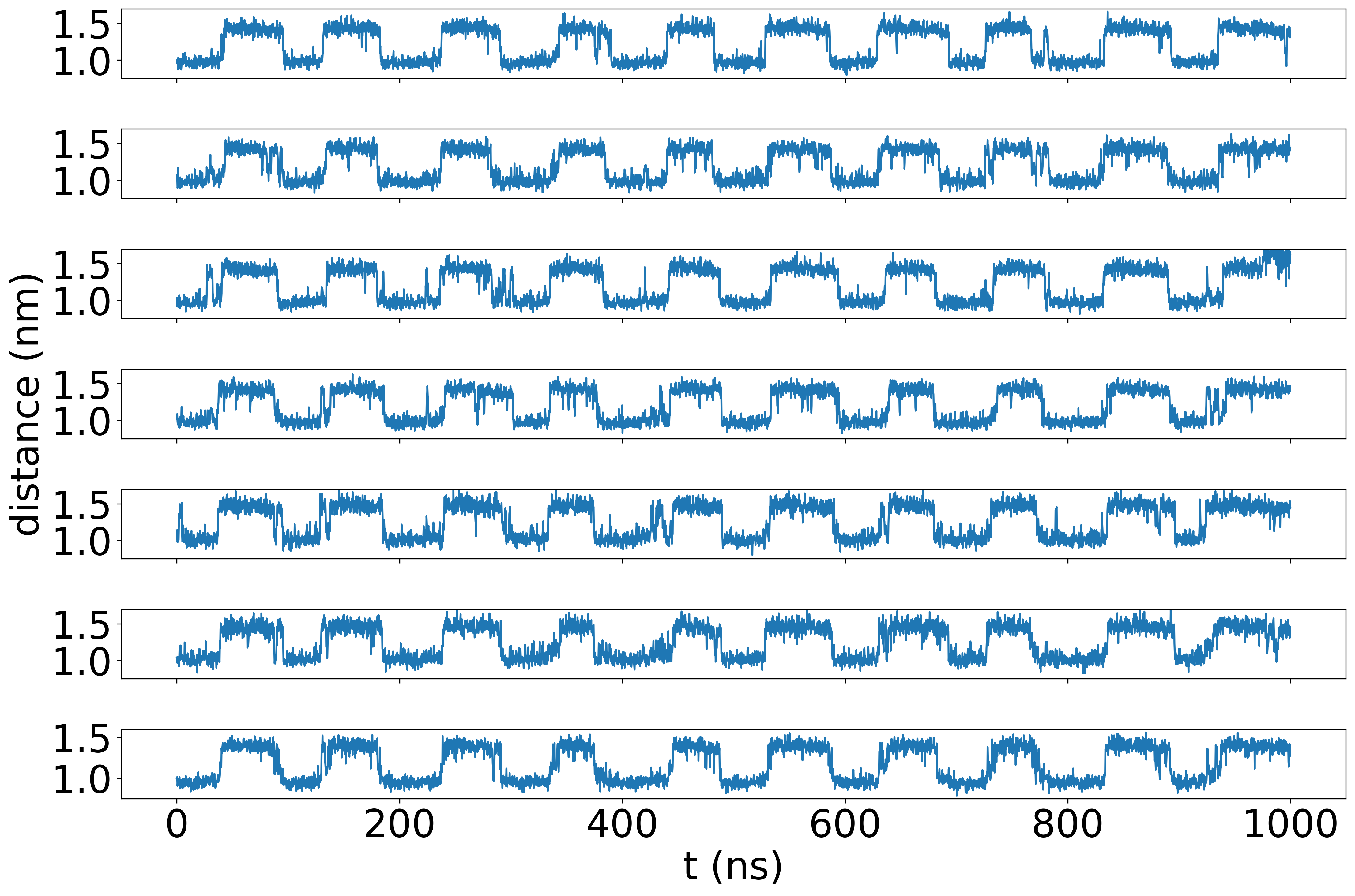}
        \label{fig:figSR_7}
        \caption{}
    \end{subfigure}
    \hfill 
    \begin{subfigure}{0.48\textwidth}
        \centering
        \includegraphics[width=\textwidth]{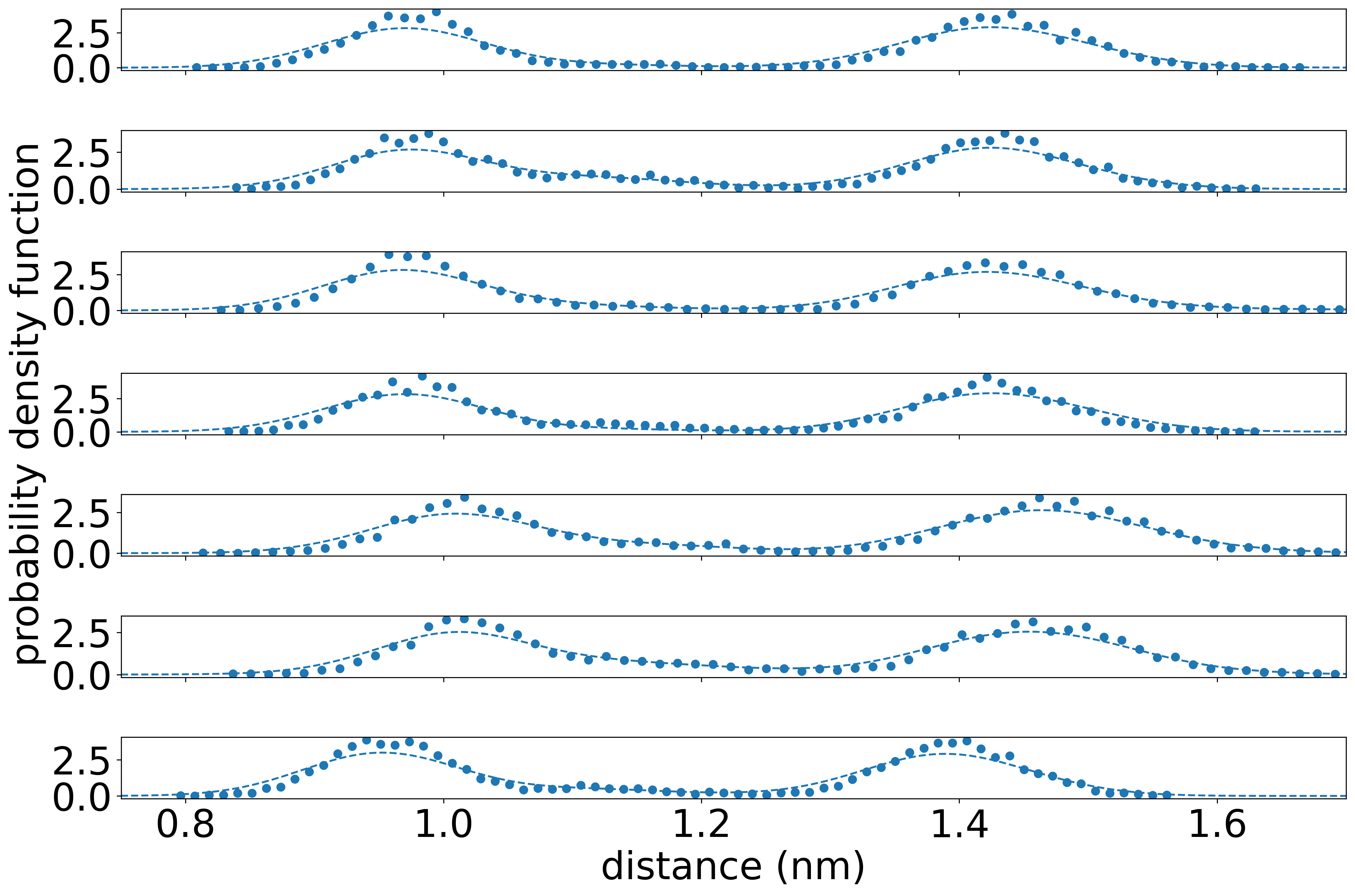}
        \caption{}
        \label{fig:figSR_pdf}
    \end{subfigure}
    \caption{Stochastic resonance of 7 oligo-PF-5 nanosprings.(a) Temporal dynamics of end-to-end distances in seven oligo-PF-5 nanosprings at oscillating field amplitude $E_{0}=\SI{0.15}{\volt\per\nano\metre}$. Each panel represents the trajectory of distance changes between the ends of one spring subjected to a symetrical tensile force \(F \approx \SI{103}{\pico\newton}\) over a 1000 ns simulation period. (b) Probability density functions for the end-to-end distance of seven oligo-PF-5 nanosprings under equilibrium conditions. Each panel represents the distribution of distances over a 1000 ns simulation period, illustrating the probability for each spring.}
    \label{fig:figSR}
\end{figure}

Figure \ref{fig:figSR} presents typical vibrations of the end-to-end distance of the oligo-PF-$5$ spring in the stochastic resonance mode, along with the power spectrum of these vibrations.

After some preliminary analysis, we decided to examine frequency response at oscillating field amplitude $E_{0}=\SI{0.15}{\volt\per\nano\metre}$, because system behavior resembled forced oscillations and not stochastic resonance (mean lifetime in state was directly proportional to external signal period). 
According to the theory of stochastic resonance \cite{gammaitoni_stochastic_1998,wellens_stochastic_2004}, the primary resonance peak was observed at a frequency of $\nu=\nicefrac{1}{2\tau}$, where the period of the applied oscillating field was equal to twice the mean lifetime of the states in the spontaneous vibration mode.

\subsection{Synchronization of Nanospring Dynamics}

To study the effects of synchronization we employed correlation matrices to analyze the trajectories of individual springs during episodes of spontaneous vibrations (SV) and stochastic resonance (SR). These matrices were constructed using Pearson correlation coefficients, which range from 0 to 1. A coefficient of 1 indicates full synchronization, where movements of springs are completely correlated, whereas a value of 0 signifies no correlation between the movements of the springs.

\begin{figure}[htbp]
    \centering
    \begin{subfigure}{0.48\textwidth}
        \centering
        \includegraphics[width=\textwidth,valign=c]{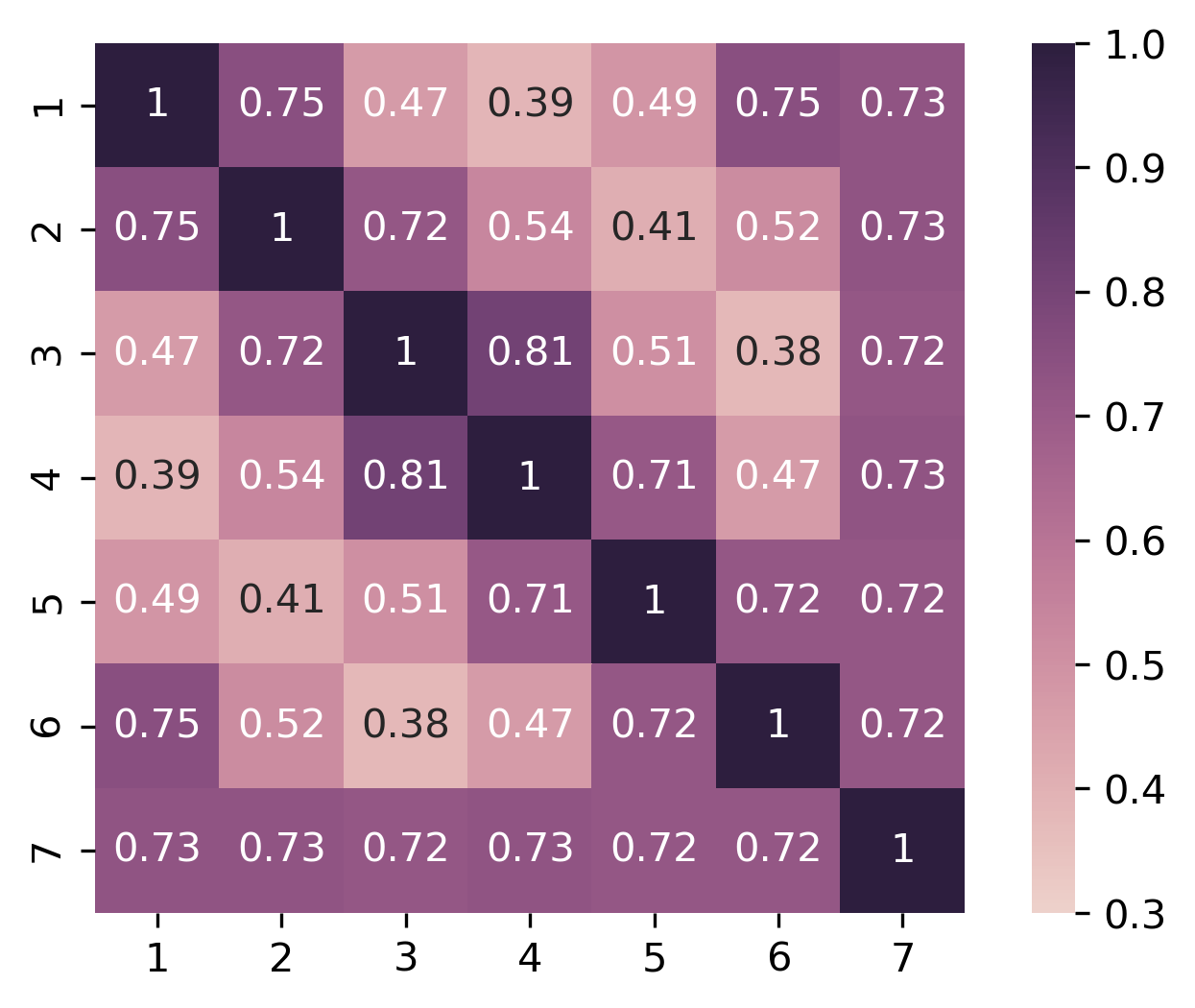}
        \label{fig:SV_corr_matrix}
        \caption{}
    \end{subfigure}
    \begin{subfigure}{0.48\textwidth}
        \centering
        \includegraphics[width=\textwidth]{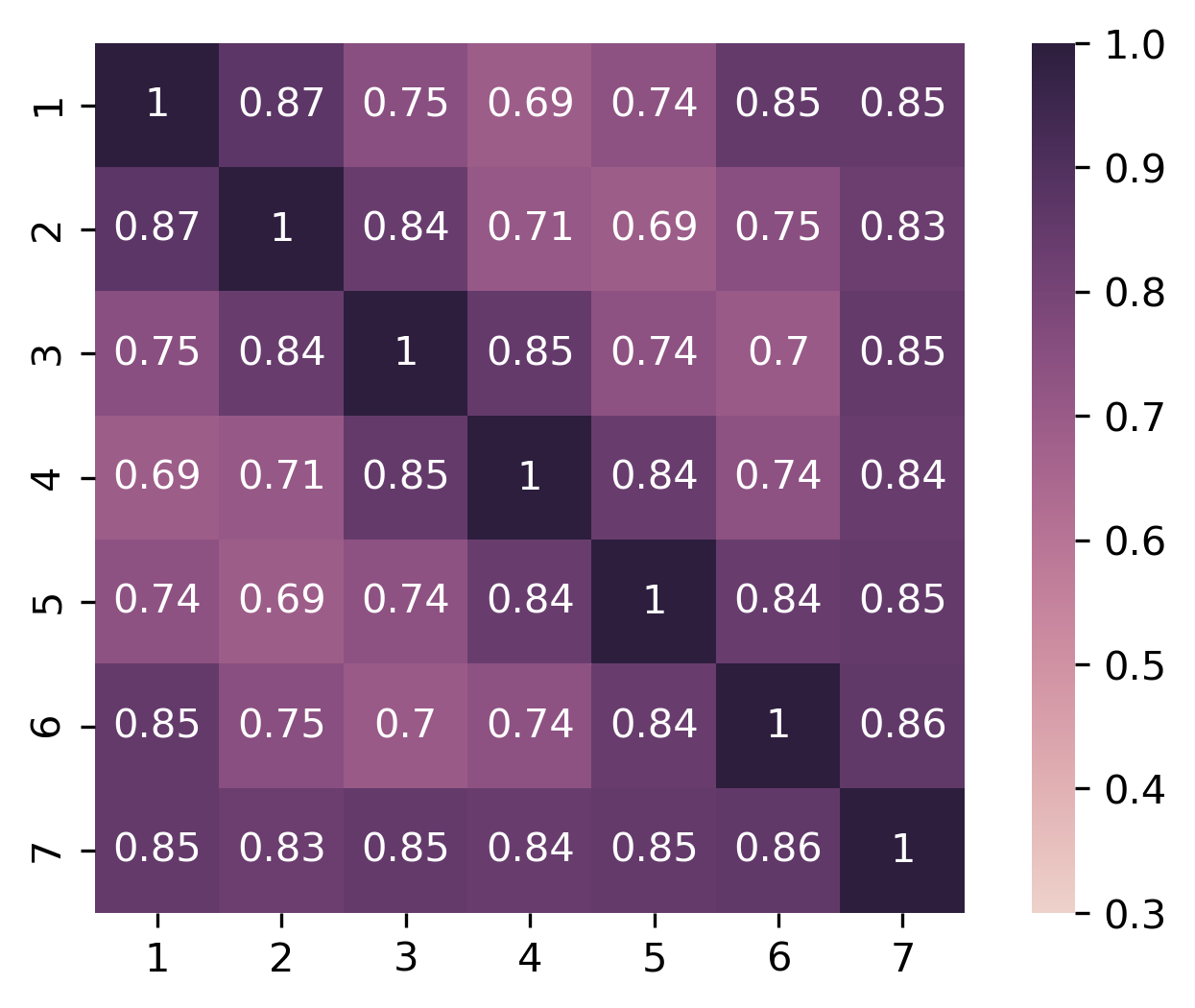}
        \caption{}
        \label{fig:SR_corr_matrix}
    \end{subfigure}
    \caption{Correlation matrix displaying Pearson correlation coefficients among the trajectories of seven oligo-PF-5 nanosprings under spontaneous vibrations (a) and stochastic resonance (b) conditions.}
    \label{fig:Corr_matrix}
\end{figure}

Figures \ref{fig:Corr_matrix}a and \ref{fig:SR_corr_matrix} display the correlation matrices for SV and SR, respectively. In the case of spontaneous vibrations, as illustrated in Figure \ref{fig:Corr_matrix}a, the matrix reveals moderate correlations, particularly among neighboring springs, with coefficients predominantly ranging from 0.44 to 0.71. This indicates a significant but not complete synchrony among the springs' movements.

On the other hand, the stochastic resonance condition, depicted in Figure \ref{fig:SR_corr_matrix}, shows higher correlation coefficients, mostly between 0.72 and 0.87, suggesting a stronger interconnected behavior and enhanced synchrony among the springs. This enhanced correlation in the stochastic resonance scenario could be attributed to the alignment of spring movements facilitated by the external noise, which aids in the synchronization of their oscillatory responses.

\subsection{Dynamic behavior of the graphene plate}\label{sec:El_cap}

\begin{figure}[htbp]
    \centering
    \begin{subfigure}{0.48\textwidth}
        \centering
        \includegraphics[width=\textwidth,valign=c]{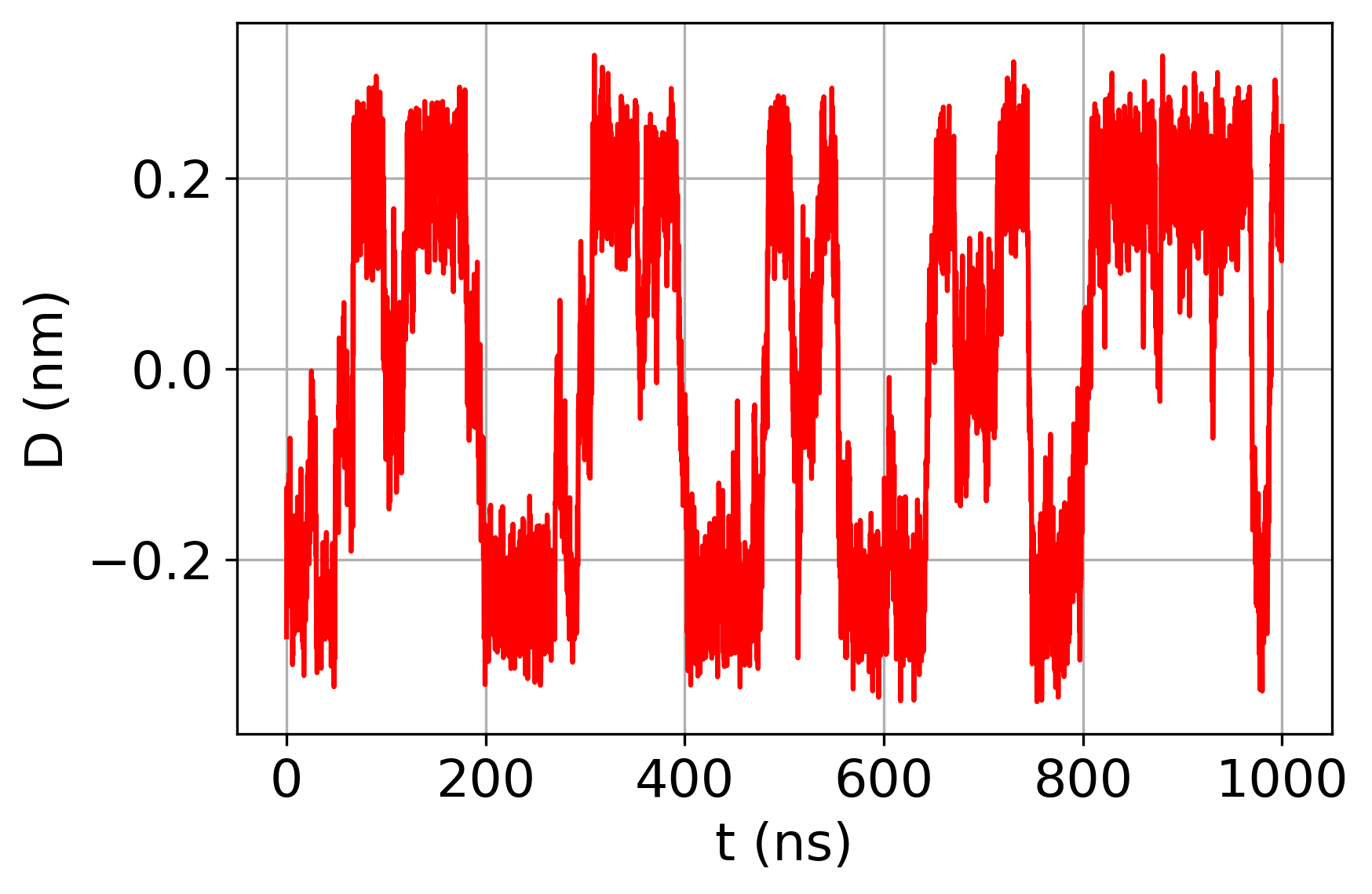} 
        \caption{}
        \label{fig:figCap_SV}
    \end{subfigure}
    \hfill 
    \begin{subfigure}{0.48\textwidth}
        \centering
        \includegraphics[width=\textwidth]{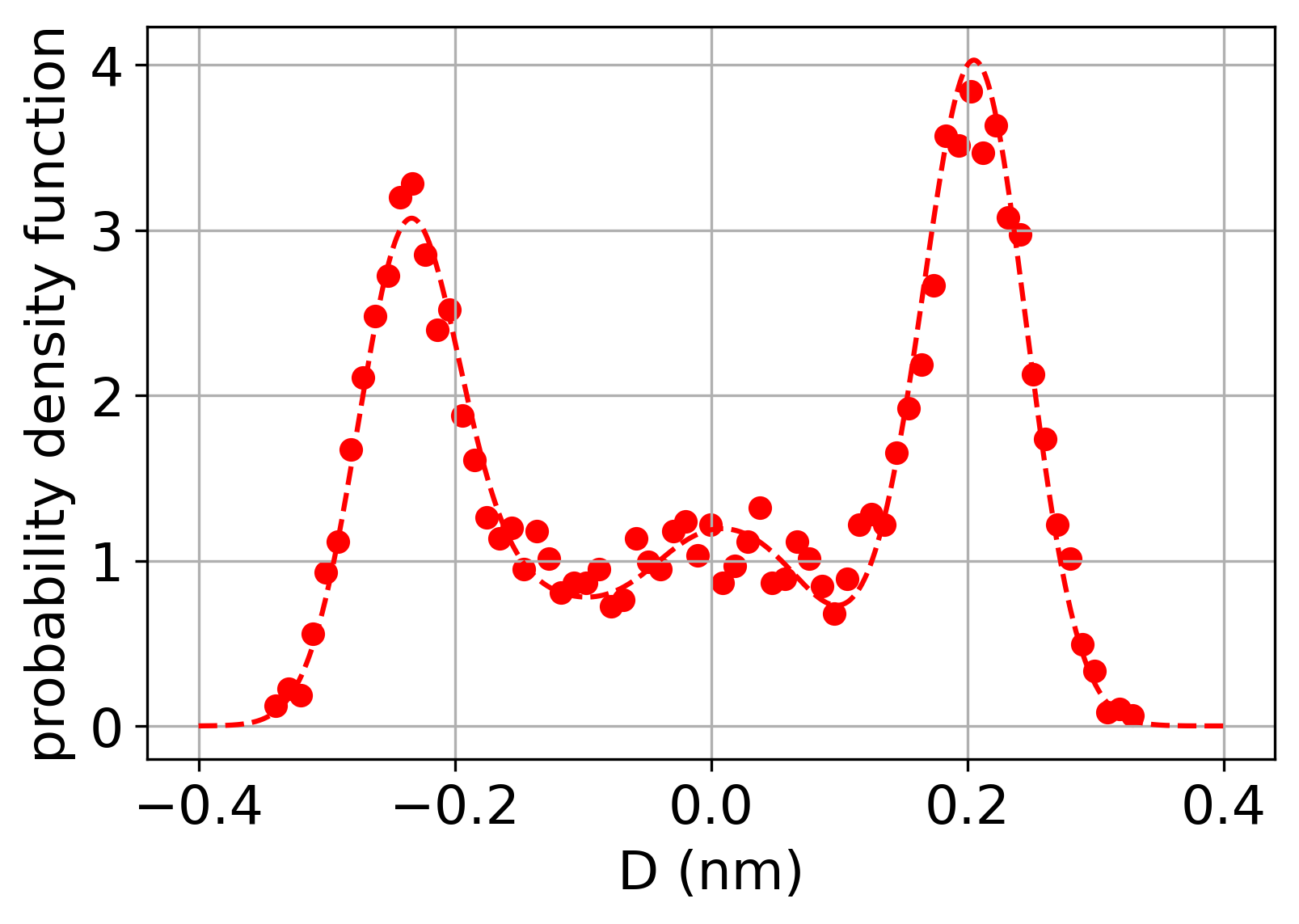} 
        \caption{}
        \label{fig:figCap_pfd_SV}
    \end{subfigure}

    \begin{subfigure}{0.48\textwidth}
        \centering
        \includegraphics[width=\textwidth,valign=c]{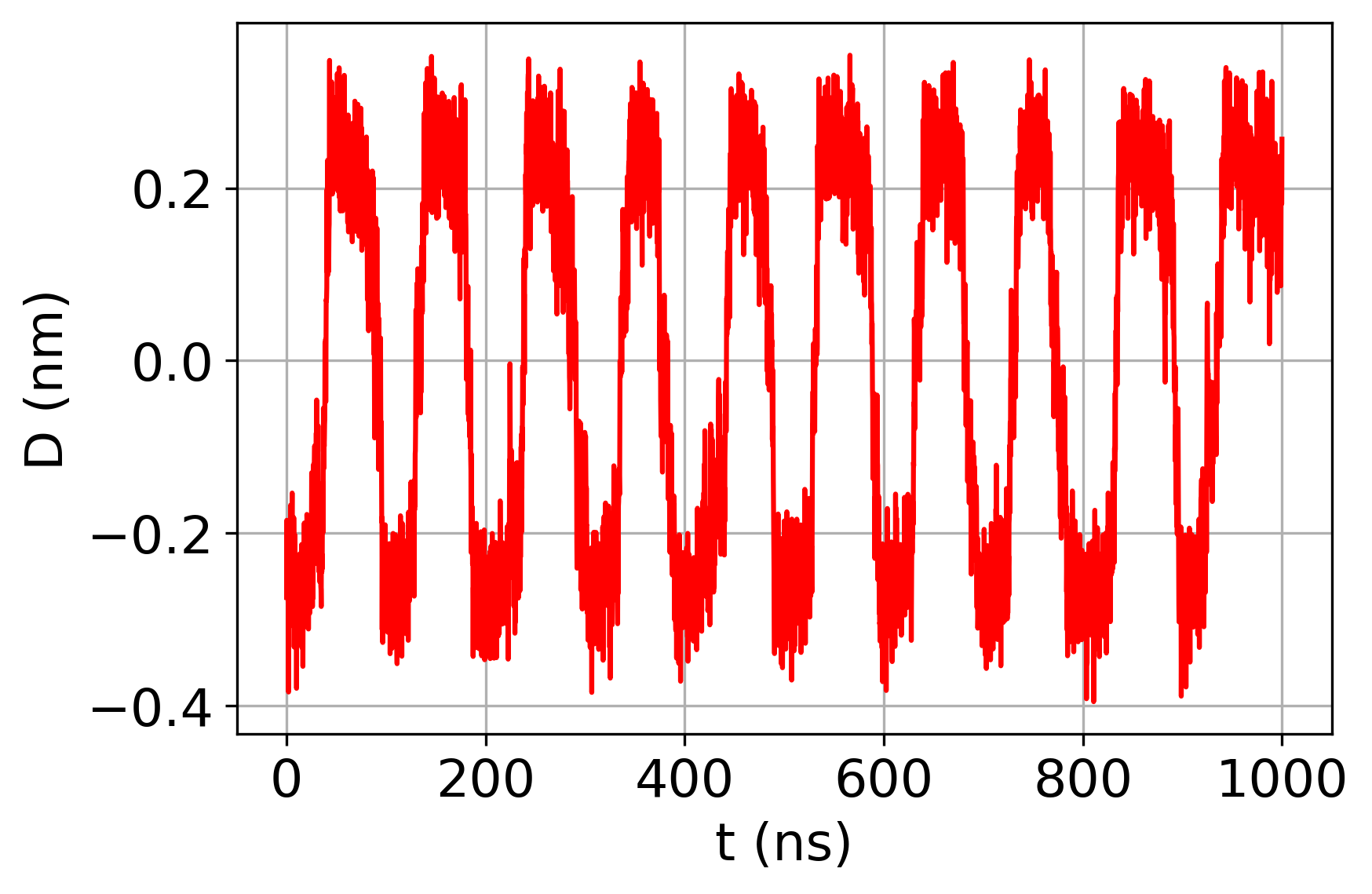} 
        \caption{}
        \label{fig:figCap_SR}
    \end{subfigure}
    \hfill 
    \begin{subfigure}{0.48\textwidth}
        \centering
        \includegraphics[width=\textwidth]{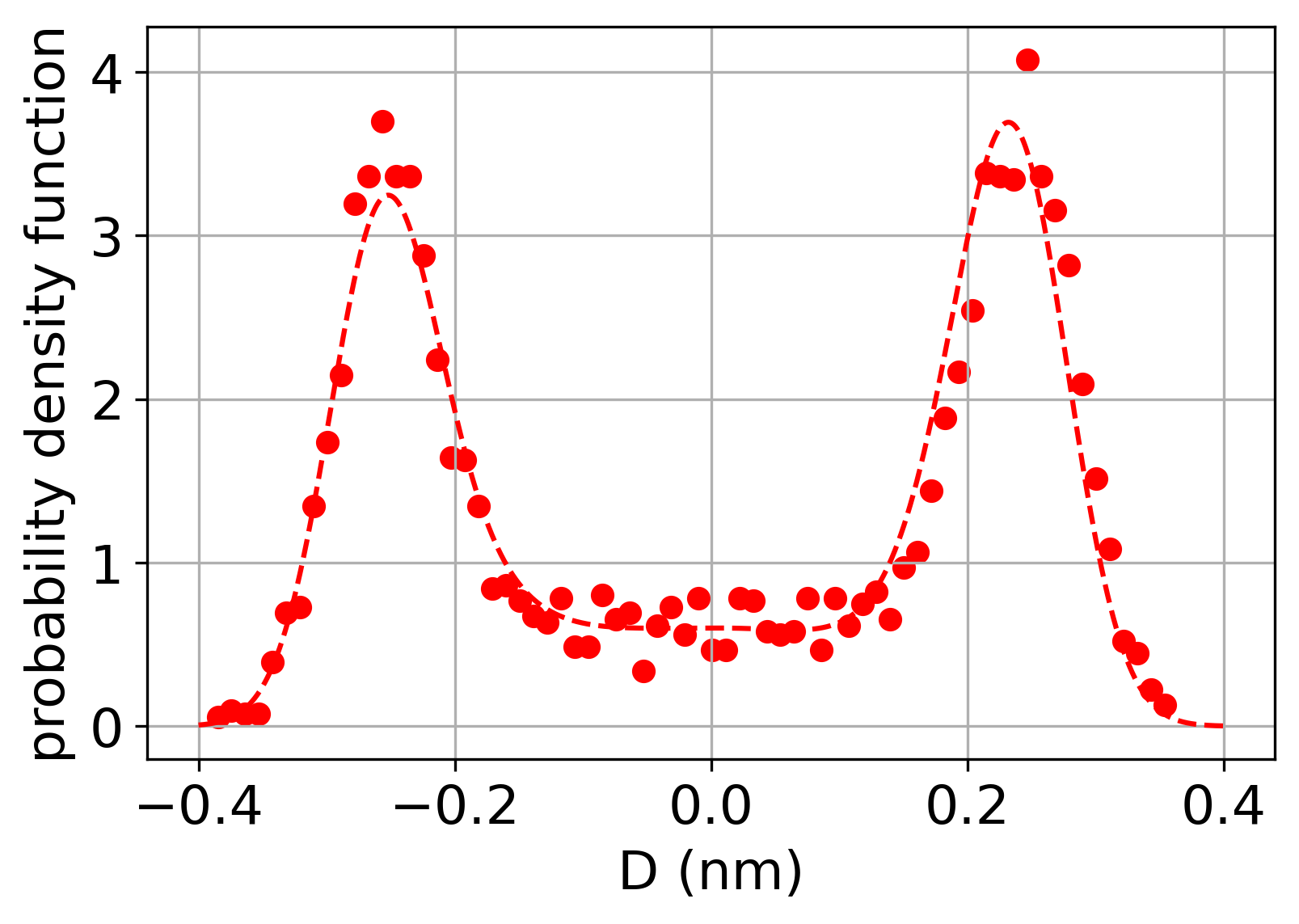} 
        \caption{}
        \label{fig:figCap_pfd_SR}
    \end{subfigure}
    
    \caption{Position of the z-coordinate of the graphene plate over time, illustrating the dynamic behavior under spontaneus vibration condition (a) and its probability density function (b). Position of the z-coordinate of the graphene plate over time, illustrating the dynamic behavior under stochastic resonanace condition (c) and its probability density function (d). All trajectories are centered relative to the time-averaged mean. Additionally, the time-averaged mean of the overall z-coordinate is shifted to zero.}
    \label{fig:figCap}
\end{figure}

To estimate the motion of the graphene plate along the z-axis, along which the tensile force vector is directed for each spring, we introduced the mean value of the z-coordinates of the carbon atoms in the graphene plate, $D$. This parameter provides a measure of the overall displacement of the graphene plate under the applied tensile forces. The resulting time dependence of $D$ for SR and SV is presented in Figures \ref{fig:figCap_SV} and \ref{fig:figCap_SR} respectively. These figures illustrate the time evolution of the mean z-coordinate, showing the vibrations of the plate caused by the vibrations of the nanosprings.

Based on the resulting trajectory, a probability density functions of SV and SR were constructed, as shown in Figures \ref{fig:figCap_pfd_SV} and \ref{fig:figCap_pfd_SR}. The probability density function allows us to analyze the dynamical states of the system, providing insights into the stability and transitions between different states. Both probability density functions of SV and SR exhibit bimodal distributions with distinct peaks, indicating preferred positions for the z-coordinate. In the SR configuration, the primary peak is around -0.2 nm and the secondary peak around 0.2 nm, both reaching a probability density of about 3, suggesting two main dynamic states. In contrast, the SV configuration not only shows similar behavior with peaks at approximately -0.25 nm and 0.25 nm but also reveals a third intermediate state. This intermediate state is characterized by a smaller peak around 0 nm, indicating a minor yet significant likelihood of the plate z-coordinate residing in this central position. These shifts and the presence of the intermediate state suggest more complex dynamic behavior of the plate in the SV configuration compared to the SR configuration.

Furthermore, the system's dynamic behavior can be harnessed in practical applications where variable properties based on the nanosprings' positions and movements are beneficial. By exploiting the distinct dynamic states and transitions, it is possible to design advanced electronic components with tunable properties, responsive to mechanical stimuli at the nanoscale. The ability to switch between different dynamic states could be particularly useful in developing devices that adjust their properties in response to external forces or environmental changes, enhancing their functionality in nanoscale applications.

\begin{figure}[ht]
    \centering
    \includegraphics[width=.5\textwidth]{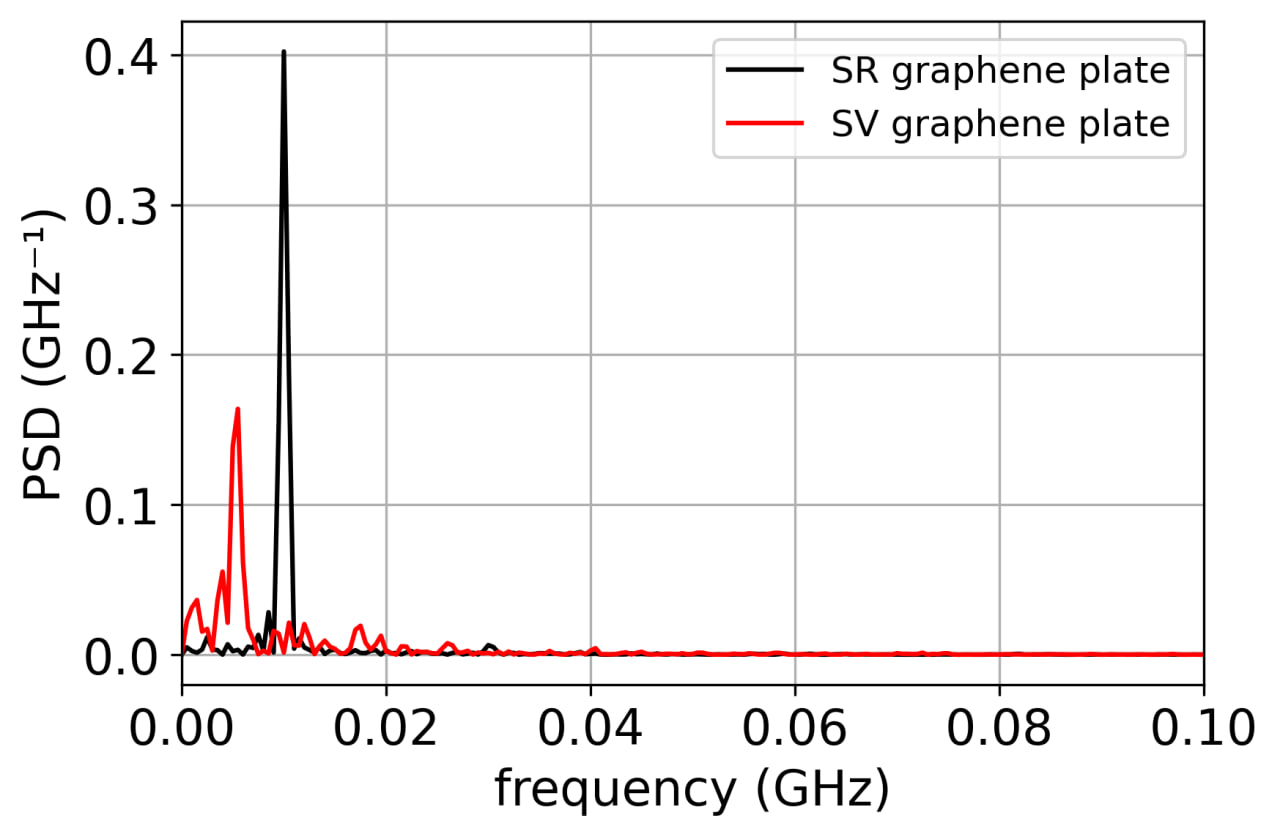}
    \label{fig:PSD_SV}
    \caption{Power spectral density (PSD) of the oscillation of the graphene plate in the cases of spontaneous vibrations (red curve) and stochastic resonance (black curve).}
    \label{fig:fig_psd_sr}
\end{figure}

Figure \ref{fig:fig_psd_sr} shows the power spectral density (PSD) of oscillations of the position of the z-coordinate of the graphene plate over time under two conditions: spontaneous vibrations (red curve) and stochastic resonance (black curve). The PSD under spontaneous vibrations displays a peak in the low-frequency region around 0.01 GHz, indicating the presence of low-frequency oscillatory modes due to thermal fluctuations. In contrast, the PSD under stochastic resonance conditions exhibits a more pronounced and well-defined peak at the frequency, corresponding to the weak oscillating signal, demonstrating the enhanced oscillatory response when driven by external periodic stimuli. This comparison highlights the amplification of weak signals through stochastic resonance, resulting in sharper and more coherent vibrational modes in the nanospring system as whole.

\subsection{Comparison of vibration spectrum of the graphene sheet and an isolated nanospring.}\label{sec:Single}

\begin{figure}[ht]
    \centering
    \includegraphics[width=.5\textwidth]{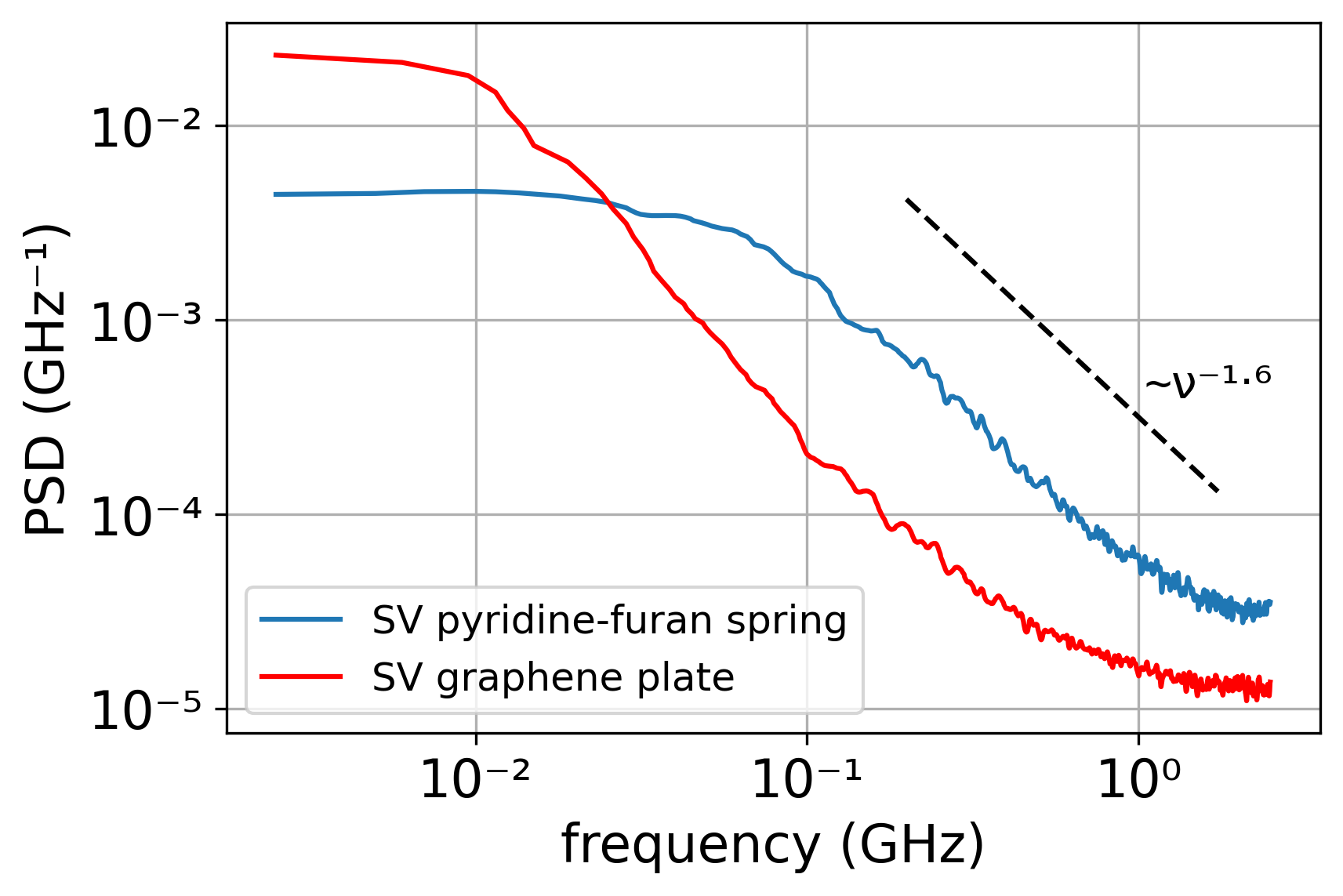}
    \label{fig:PSD_SV}
    \caption{Power spectral density (PSD) of a single pyridine-furan nanospring compared to the system of seven pyridine-furan nanosprings. Dashed line shows the slope coefficient, representing the power law of PSD decay.}
    \label{fig:fig4}
\end{figure}

Figure \ref{fig:fig4} shows the power spectral density (PSD) of SV of a single pyridine-furan nanospring compared to a system of seven pyridine-furan nanosprings. The PSD of the single nanospring (blue curve) shows a relatively stable power density across the frequency range, with a gradual decrease as the frequency approaches 0 GHz. This pattern is typical for a single oscillator with well-defined vibrational modes.

In contrast, the PSD of the system of seven nanosprings (orange curve) exhibits a more significant decline in power density as the frequency decreases, suggesting a shift to lower frequencies. This shift indicates that the coupled system exhibits collective behavior, which modifies the frequency response compared to the single nanospring. The collective dynamics in the system of seven nanosprings lead to enhanced low-frequency oscillations, highlighting the impact of coupling on the overall vibrational characteristics of the nanosprings. Therefore the system of seven nanosprings could be interprated as low-frequency filter.

Furthermore, the presence of these low-frequency peaks in the PSD of the coupled system highlights the role of the graphene plate in mediating interactions between the nanosprings, allowing them to exhibit collective behavior distinct from the behavior of an isolated spring. This collective vibrational behavior has important implications for the design of nanoscale devices, as it can lead to enhanced sensitivity and functionality by leveraging the synchronized dynamics of multiple nanosprings.

The differences in the PSDs between the single and coupled nanospring systems underscore the significance of nanospring coupling in influencing dynamic behavior. The observed enhanced low-frequency oscillations and prolonged coherence times in the coupled system suggest that such designs could be advantageous for applications requiring stable and synchronized responses to external stimuli. These findings provide a deeper understanding of how nanospring interactions can be exploited to optimize the performance of nanoscale mechanical systems.

\section{Discussion}\label{sec:Dis}

The main finding of this work is that spontaneous vibrations and stochastic resonance are not exclusive to single oligo-PF springs \cite{avetisov2021short}, but are also present in more complex systems. However, the bistability effects might vary for different systems, and such variations are discussed below.

Under a tensile force of \( F = 103 \) pN, spontaneous vibrations were observed in the oligomers, exhibiting symmetric distributions of the state visits for each spring. Analysis of correlation coefficients between pairs of springs revealed significant synchronization, particularly between neighboring springs. Moreover, synchronization of the dynamics among the seven oligomers was observed at low-frequency components of the thermal noise, indicating collective behavior in the system. In the case of seven oligomers, SV were more pronounced compared to individual oligomers, suggesting potential cooperative dynamics.

From the perspective of nonlinear dynamical systems, the dynamics of the oligo-PF-5 spring bifurcate at the critical force \( F_s = 103 \) pN. Beyond this critical tensile load, the spring becomes bistable and exhibits spontaneous vibrations, alternating between the squeezed and stress-strain states. The average end-to-end distances of the spring in the squeezed and stress-strain states differ by approximately 0.35 nm, allowing for a clear distinction between these two states.

The first note concerns the implications of the observed synchronization effects for practical applications. The enhanced synchronization among multiple nanosprings could lead to improved signal processing capabilities in nanoscale devices. Such devices could benefit from the collective bistability, allowing for more robust and sensitive responses to external stimuli.

The second note pertains to the potential for tuning the bistability characteristics of the nanosprings through modifications in their molecular structure or the external environment. By adjusting factors such as chemical structure of the oligomers, the type of solvent, or the strength of the applied tensile force, it may be possible to fine-tune the bistable behavior and optimize the performance of nanoscale devices.

The next note addresses the scalability of the hybrid system design. While the current study focuses on a system with seven nanosprings, future work could explore the behavior of larger arrays of nanosprings. Understanding how the collective dynamics scale with the number of nanosprings could provide valuable insights for designing larger, more complex nanoscale systems. The theoretical sections explains that AWT is growing with growing of the system size (number of coupled units) and coupling strength. Therefore for larger system the coupling strength could be decreased to rich meaningful AWT of the system vibrations. 

The last note concerns the stochastic resonance observed in the system. The ability to harness stochastic resonance for enhancing weak signal detection opens new avenues for developing advanced materials and devices. By leveraging the interplay between bistable dynamics and stochastic perturbations, it may be possible to create nanoscale systems with unprecedented sensitivity and functionality. Furthermore, the observed dynamic behavior can be harnessed in practical applications such as nanomechanical and nanoelectromechanical operational units. This could enable the development of advanced electronic components with tunable properties, responsive to mechanical stimuli at the nanoscale. 

\section{Conclusions}\label{sec:Con}

Stochastic resonance, emerging from the interaction between bistable dynamics and stochastic perturbations, significantly enhances weak signal detection in various systems, including nanoscale devices like oligo-PF-5 nanosprings. Our study demonstrates that spontaneous vibrations and stochastic resonance are not only present in single oligo-PF-5 springs but also in more complex systems of multiple nanosprings. These nanosprings exhibit significant correlations and enhanced synchronization under tensile force and low-amplitude periodic electric fields, suggesting potential cooperative dynamics for improved signal processing.

By tuning the bistability characteristics through modifications in molecular structure or external environment, it is possible to optimize the performance of these nanoscale devices. Factors such as chemical structure of the oligomers, solvent type, and tensile force can be adjusted to fine-tune the bistable behavior, providing a versatile platform for advanced materials with specific dynamic responses.

The scalability of the hybrid system design is promising. While this study focused on seven nanosprings, future research could explore larger arrays to understand the collective dynamics at a larger scale. Insights from such studies could inform the design of more complex nanoscale systems with enhanced functionalities.

In conclusion, the observed collective bistability and synchronization of nanosprings highlight the potential for developing responsive nanoscale devices. Further investigation into the scalability and tunability of these systems could lead to significant advancements in nanotechnology and materials science, opening new avenues for innovative applications.

\section*{Funding}
The authors acknowledge the financial support of the Design Center for Molecular Machines, Moscow, Russia, under the Business Contract № 202-24-5.05.2024 with N. N. Semenov Federal Research Center for Chemical Physics, Russian Academy of Sciences.

\section*{Acknowledgments}
The research is carried out using the equipment of the shared research facilities of HPC computing resources at Lomonosov Moscow State University. We also thank Vladimir Bochenkov for helpful discussions.


\end{document}


\section{Simulation protocol}
\subsection{Parameters for Molecular Dynamics simulation\label{SI1}}
Morphology simulations were performed using the Gromacs2019 \cite{abr2015} simulation package. Lennard-Jones parameters were taken from the OPLS-AA \cite{kam2001} force field with a scaling factor of 0.5 for the 1–4 interactions. Long-range electrostatic interactions were treated using a smooth particle mesh Ewald technique \cite{ess1995} with a cut-off of \SI{1.2}{\nano\metre}. Bond vibrations were constrained with a LINCS \cite{hes2008} algorithm. All calculations were performed in the NVT ensemble using the canonical velocity-rescaling thermostat, as implemented in the Gromacs2019 simulation package.

\begin{figure}[H]
    \includegraphics[width=\linewidth]{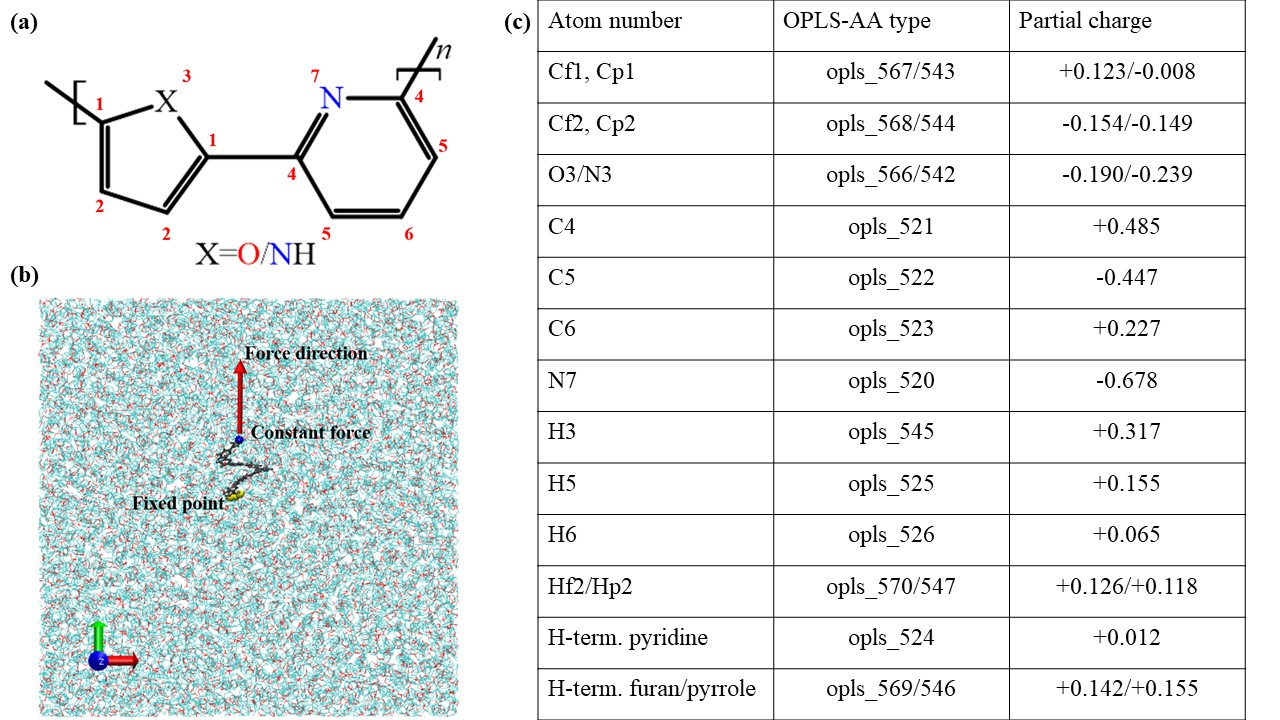}
    \centering
\caption{(a) The chemical structure of a PF/PP monomer, (b)a scheme of the simulation box for the oligo-PF5 system, (c) parameterization of PF/PP oligomers in OPLS-AA force field types and partial charges.}
\label{fig:S1}
\end{figure}

The parameterization of oligo-PP springs is presented in Figure \ref{fig:S1}a-b. The OPLS-AA force field contains parameters only for single pyridine and furan molecules. First, C4 and C6 atoms have a covalent bond instead of hydrogen atoms, and the partial charge of hydrogens was added to carbon atoms to maintain neutrality. All the other partial charges are taken from the OPLS-AA force field. Second, there is only one undefined angle bond for atoms C4-C6-N11, which is the CW-CA-NC angle type in OPLS-AA. We used the CA-CA-NC angle due to the similarity of CA and CW types. Both correspond to aromatic sp2-hybridized carbon.

The simulation was started from a random initial configuration. An equilibrated state was reached in two steps. First, a short run of \SI{10}{\pico\second} was carried out in the NVT ensemble with a time step of \SI{0.01}{\femto\second} using the leapfrog integrator for motion equations. Then the second part of equilibration was done for another \SI{10}{\nano\second} with a time step of \SI{2}{\femto\second}. After equilibration, a long trajectory of \SI{500}{\nano\second} was obtained to achieve a clear picture of transitional behavior.

To simulate the behavior of the oligo-PP5 under the external load, one end of it was fixed and the other end was pulled by an external force (see Figure \ref{fig:S1}c). We used a simulation box of size $7 \times 7 \times 7$ \si{\nano\metre\cubed} due to technical aspects of the Gromacs pulling algorithm. The axis of the initial conformation of oligo-PP5 was oriented in the XY plane. The longitudinal load F was applied along the Z-direction to the center of mass of the last monomer unit and directed toward the attraction point, which was located along the vector connecting the left and right ends of the molecule.

In both cases, the center of mass of the first monomeric unit was fixed using a spring potential of $k = \SI{100}{\kilo\joule\per\mole\per\nano\metre\squared}$; other specific constraints for bond length or atom positions were applied.

\begin{figure}[ht]
\centering
\includegraphics[width=.2\textwidth]{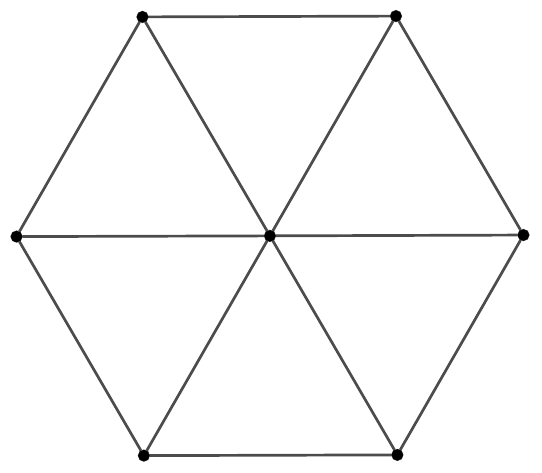}
\caption{Graph representation of the adjacency matrix $A$ determining the topology of interparticle interactions.}
\label{Fig_adj}
\end{figure}

The figure illustrates the power spectral density (PSD) for each pyridine-furan nanospring in a construction of seven nanosprings, showing data for both spontaneous vibrations (red curves) and stochastic resonance (black curves). The PSD under spontaneous vibrations exhibits distinct peaks in the low-frequency region, typically around 0.01 GHz, indicating the presence of low-frequency oscillatory modes due to random thermal fluctuations. In contrast, the PSD under stochastic resonance conditions features more pronounced and well-defined peaks concentrated around slightly higher frequencies, highlighting the enhanced oscillatory response of the nanosprings when driven by external periodic stimuli. The comparison between the red and black curves demonstrates that stochastic resonance amplifies weak signals, resulting in sharper and more coherent peaks in the PSD, which is indicative of the collective behavior and synchronization among the nanosprings facilitated by the external driving force. 

\begin{figure}[ht]
\centering
\includegraphics[width=.9\textwidth]{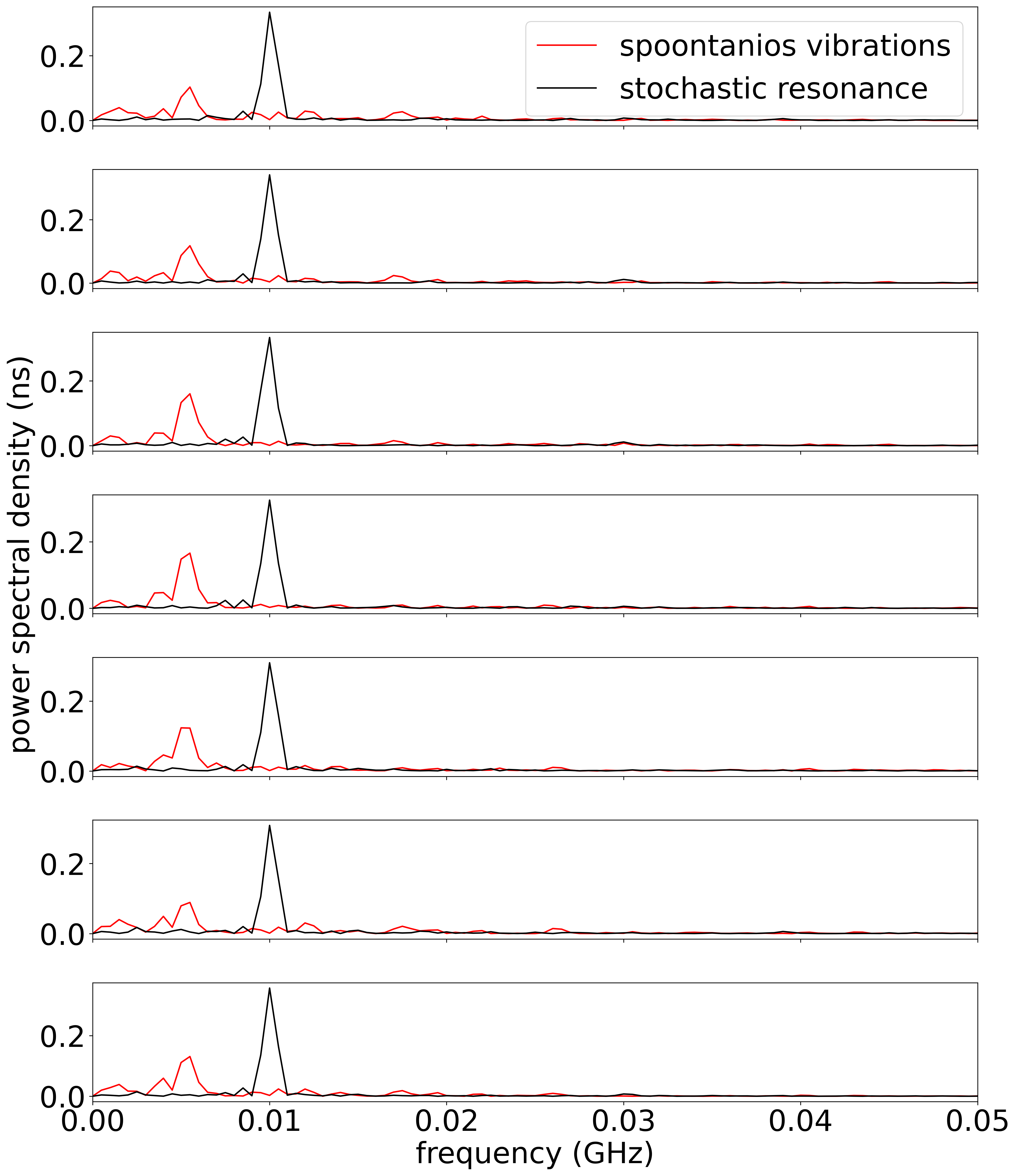}
\caption{Power spectral density (PSD) for each nanospring in the construction during spontaneous vibrations (red curves) and stochastic resonance (black curves).}
\label{Fig_res}
\end{figure}

\subsection{Oligo-PP-$5$ in water}
In order to find out the impact of THF on the dynamics of oligo-PP, we also model its stretching in water. The simulation protocol in this case is similar to the one described in the section Parameters for Molecular Dynamics simulation above. One end of the oligo-PP$5$ was fixed and the pulling force was applied to another end. The end-to-end ($R_{e}$) distance both in water gradually increases along with the force up to the value of $F$ = \SI{305}{\pico\newton} at which point the oligo-PP$5$ helix turns to the fully stretched state
. Note that the oligo-PP$5$ in water is present in the squeezed state at the pulling forces below $F$ = \SI{305}{\pico\newton} due to the hydrophilic nature of the solvent.


\subsection{Parameters for periodic signal}\label{SI:2}
Stochastic resonance was obtained by applying a periodic signal leading to the swinging of the oligo-PF/PP bistable potential. An oscillating force was implemented by setting a charge ($-1$) on the oligomer, adding a compensating charge ($+1$) as a counter ion in the solvent and applying a periodical electric field. The additional charge was placed on the end group of the oligo-PF-$5$/oligo-PP-$5$ and spread between the first, second and fifth atoms of the furan/pyrrole group (see Fig. \ref{fig:S1}a). The additional charges are equally distributed between the chosen atoms.

The periodic field in the Gromacs2019 package is defined by an equation $$ E(t) = E_0 \exp{\left[-\frac{(t-t_0)^2}{2\sigma^2}\right]}\cos{\omega(t-t_0)},$$ where the exponential part modulates the periodic part with the pulsing behaviour, $E_0$ is the amplitude of the signal and $\omega t_0$ is the oscillation phase. Here we use only the static part when $\sigma=0$ along y-axes (see Figure \ref{fig:S1}b). An external oscillating electrical field was directed along the constant force direction in the oligo-PF-$5$/oligo-PP-$5$ case.

\subsection{Influece of the charges}\label{SI:3}

The SR regime of the both oligo-PP-$5$ and oligo-PF-$5$ springs was examined by applying an additional oscillating electrical force on the unit charge preset on the pulling end of the spring while a compensative charge was placed $\sim \SI{2.2}{\nano\metre}$ "below" the spring (if pulling direction is "up"). However, such a modification changed drastically the bistable behaviour of the springs: even with $\vec{F} = 0$ the bistability was observed with spontaneous vibrations present, while the equilibrium value of $F_{eq}$ (when mean time spent in open and closed states is approximately equal) for oligo-PP-$5$ was only $  \SI{3.5}{\pico\newton} $. This result indicates that an artificial negative charge decreases the size of the closed state energy minimum, presumably because of its interactions with partial charges on the aromatic ring that comes in contact with the charged end of the spring in closed state. We also observed that positively charged end of the spring influences the $F_{eq}$ in opposite way (it increases compared to the non-charged case), which partially confirms the hypothesis about interactions with partial charges.

\subsection{Torsion angle dynamics due to vibrations}\label{SI:4}
\begin{figure}[H]
    \includegraphics[width=\linewidth]{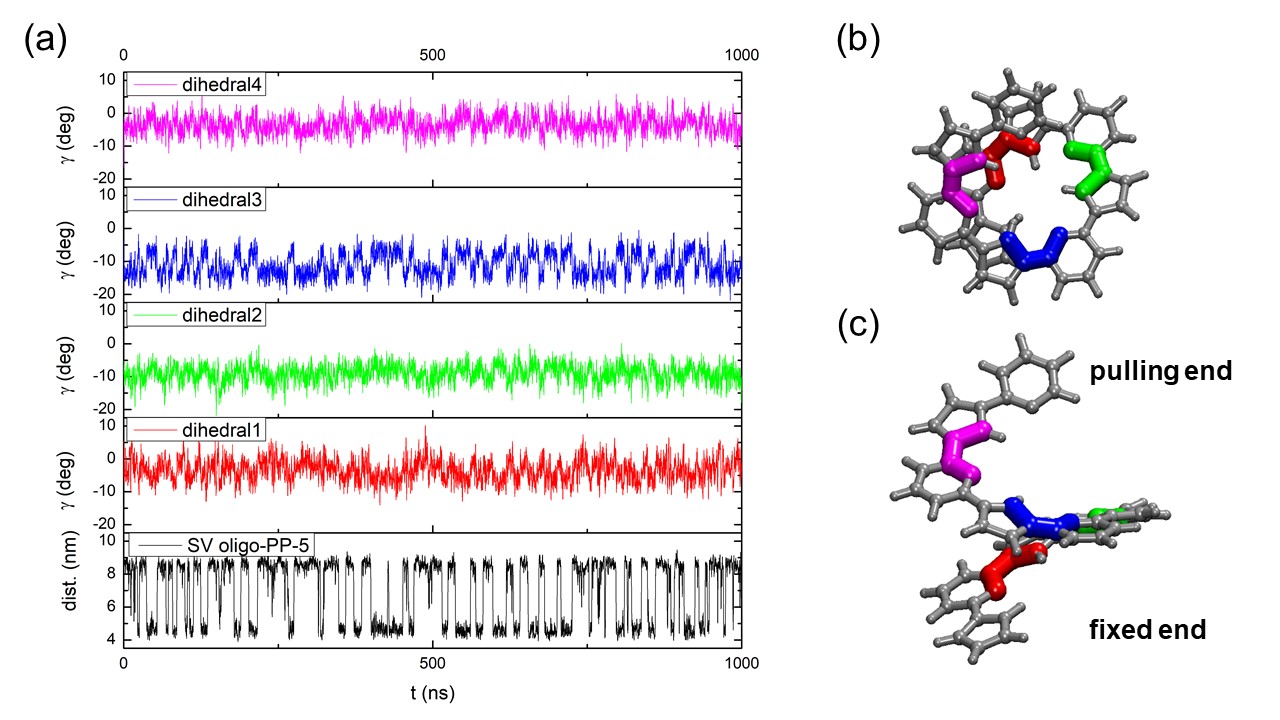}
    \centering
\caption{ (a) Trajectories for the oligo-PP-5 spontaneous vibrations and N-C-C-N dihedrals along the spring structure. (b) Top and (c) side views of the oligo-PP-5 with selected dihedral groups.}
\label{fig:S4}
\end{figure}
We carried out a detailed analysis of torsion angles in the oligo-PP-5 spring during spontaneous vibrations. We chose the N-C-C-N dihedral as the most representative case since it connects different monomers. A distinguishable bistable trajectory is observed only for the dihedral, which bonds the 3rd and 4th monomers. The correlation coefficient for the trajectory of the dihedral torsion angle and the trajectory of oligo-PP-5 spontaneous vibrations was 0.86 (Figure \ref{fig:S5}a, blue curve). The other correlation coefficients had smaller values since the other monomers are closer to the oligo-PP-5 spring ends, which are under external restrictions such as fixing or force application. 

\subsection{Mean lifetime estimation}
\begin{figure}[H]
    \includegraphics[width=\linewidth]{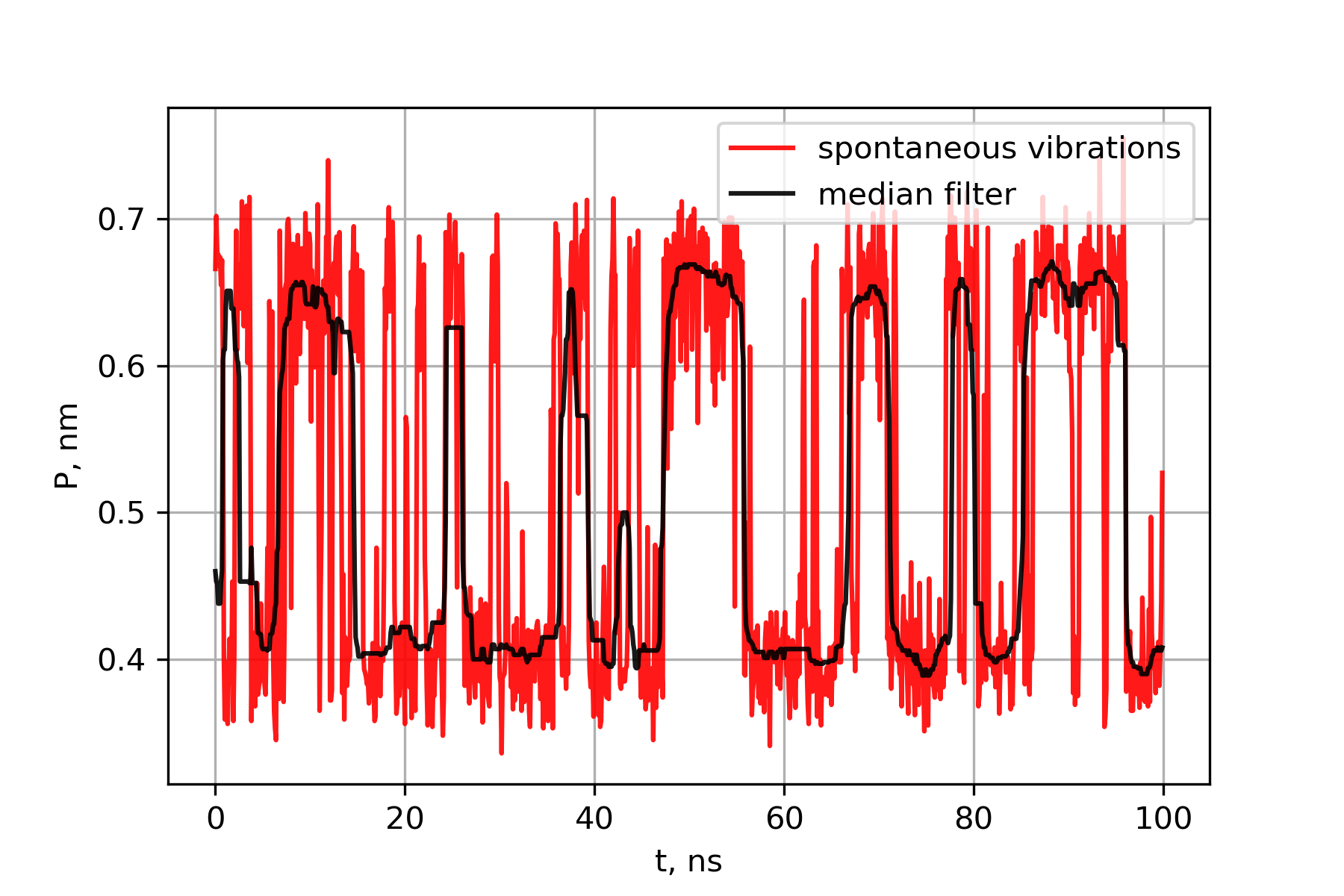}
    \centering
\caption{ The spontaneous vibrations trajectory with and without the median filter usage.}
\label{fig:S5}
\end{figure}
In the oligo-PF/PP systems, one can observe the dynamics at two temporal scales: local and global. The first one is characterized by fast fluctuations near the squeezed/stress-strain states in case. However, the global dynamics (the switching between the ends) is more important for stochastic resonance. To calculate the mean lifetime, the local dynamics were excluded using the median filter $ndimage.median\_filter$ from an open-source software SciPy for Python 3 with a median filter window size of \SI{3.5}{\nano\second}, which corresponds to local vibrations. The resulting trajectory is shown in \ref{fig:S5}. According to the estimation, the mean lifetime for oligo-PF$5$ was $\tau=\SI{6.14}{\nano\second}$ and corresponded to the external force $F=\SI{279}{\pico\newton}$.

\bibliography{references}